**Accurate Liability Estimation Improves Power in Ascertained Case Control Studies**

Omer Weissbrod[1], Christoph Lippert[2], Dan Geiger[3] and David Heckerman[4]

[1]Computer Science Department, Technion - Israel Institute of Technology, Haifa, Israel

[2]eScience Group, Microsoft Research, Los Angeles, USA

[3]Computer Science Department, Technion - Israel Institute of Technology, Haifa, Israel

[4]eScience Group, Microsoft Research, Los Angeles, USA

Corresponding Author: David Heckerman

# Abstract

Linear mixed models (LMMs) have emerged as the method of choice for confounded genome-wide association studies. However, the performance of LMMs in non-randomly ascertained case-control studies deteriorates with increasing sample size. We propose a framework called LEAP (Liability Estimator As a Phenotype; https://github.com/omerwe/LEAP) that tests for association with estimated latent values corresponding to severity of phenotype, and demonstrate that this can lead to a substantial power increase.

# Main Text

In recent years, genome-wide association studies (GWAS) have uncovered thousands of risk variants for genetic traits[1]. Only a small fraction of disease variance is explained by discovered variants, possibly because contemporary sample sizes are relatively small and that causal variants tend to have small effect sizes[2]. To identify such variants, future studies will need to include hundreds of thousands of individuals.

Population structure and family relatedness[3] lead to spurious results and increased type I error rate. As sample sizes continue to increase, this difficulty becomes even more severe, because larger samples are more likely to include individuals with a different genetic ancestry, or related individuals.

Recently, LMMs have emerged as the method of choice for GWAS, due to their robustness to diverse sources of confounding[3]. LMMs gain resilience to confounding by testing for association conditioned on pairwise kinship coefficients between study subjects. Although designed for continuous phenotypes, LMMs have been successfully used in several large case-control GWAS[4-6], because alternative methods cannot capture diverse sources of confounding[3].

However, LMMs in ascertained case-control studies, wherein cases are oversampled relative to the disease prevalence, lose power with increasing sample size compared to alternative methods[7]. This loss is due to several model violations: Dependence between tested causal variants and variants used to estimate kinship, dependence between genetic and environmental effects, and use of a non-continuous trait (**Supplementary Note**). Thus, the use of LMMs resolves the difficulty of sensitivity to confounding, but leads to a different difficulty instead.

A possible remedy is to test for associations with a model that directly represents the case-control phenotype and takes the ascertainment scheme into account (Supplementary Note). Such models assume that observed case-control phenotypes are generated by an unobserved stochastic process with a well-defined distribution. One prominent example is the liability threshold model[8], which associates individuals with a latent normally distributed variable called the *liability*, such that cases are individuals whose liability exceeds a given cutoff. Despite their elegance, such models are extremely computationally expensive, rendering whole genome association tests infeasible in most circumstances.

As an alternative, we propose approximating such models by first estimating latent liability values and model parameters conditional on phenotypes, genotypes and disease prevalence, and then testing for association with the estimated liabilities via an LMM (Online Methods). LEAP is motivated by the observation that cases of rare diseases have a sharply peaked liability distribution (**Supplementary Fig. 1**), leading to highly accurate liability estimation (**Supplementary Note**). When testing for association in ascertained case-control studies, LEAP yields substantially increased power over naïve LMMs while remaining resilient to confounding because it largely compensates for the violations listed above.

LEAP bears similarities to several recent methods for estimating the portion of the liability that is explained by a small set of explanatory variables[9, 10]. Unlike these methods, LEAP estimates liabilities using the entire genome (**Supplementary Note**). In parallel work, Hayeck *et al.* proposed another framework called LTMLM (liability threshold mixed linear model) for association testing in ascertained case-control studies[11]. Both LTMLM and LEAP first estimate latent liability values and then test for association with these estimates. However, LTMLM tests for association with the posterior mean of the liabilities in a score test framework, whereas LEAP tests for association with a maximum a posteriori (MAP) estimate, which is often more robust to model violations and can be evaluated at a substantially reduced computational cost (Online Methods and **Supplementary Note**).

We evaluated the performance of LEAP on synthetic and real data sets, using the following methods for comparison: (i) LEAP, (ii) a standard LMM, (iii) a linear regression test using ten principal component (PC) covariates[12] (denoted Linreg+PCs), and (iv) a univariate linear regression test (Linreg) without PC covariates, used as a baseline measure. Linreg+PCs and Linreg use the linear link function to prevent evaluation bias due to using a different link function. Experiments using logistic regression yielded results very similar to those with Linreg (data not shown).

Sensitivity to confounding was evaluated by measuring type I error rates for synthetic data sets (Supplementary Note). We evaluated various combinations of population structure (quantified via the $F_{ST}$ measure[13]) and family relatedness (measured via the fraction of sibling pairs), using $F_{ST}$ levels of 0, 0.01 and 0.05, and sib-pair fractions of 0%, 3% and 30%. Only LEAP and a standard LMM properly controlled for type I error in the presence of confounding (**Supplementary Figs. 2** and **3** and **Supplementary Table 1**).

The power of the methods was evaluated according to the distribution of test statistics of causal variants[9, 10]—normalized according to the type I error rate—to prevent Linreg and Linreg+PCs from falsely appearing to be more powerful than other methods due to inflation of P values (**Supplementary Note**).

To investigate the effects of sample size and ascertainment on power, we generated ascertained case-control data sets with prevalence of 0.1%, 1% and 10%, and sample sizes in the range 2,000-10,000. The advantage of LEAP over a standard LMM increased with sample size and with decreasing prevalence (**Fig. 1** and **Supplementary Figs. 4** and **5**). In simulated samples with 0.1% prevalence and 10,000 individuals, LEAP gained an average increase of over 20% in test statistics of causal variants (**Fig. 1**), and a power increase of over 5% for significance thresholds smaller than $5 \times 10^{-5}$ (**Supplementary Fig. 4**). LEAP also outperformed other methods under more complex ascertainment schemes (**Supplementary Note**). We further verified that the increased power of LEAP stems from its accurate liability estimation in the presence of ascertainment (**Supplementary Fig. 6** and **Supplementary Note**).

Accurate liability estimation depends on the fraction of liability variance that is driven by genetic factors, called the narrow-sense heritability[2]. A higher heritability is expected to improve estimation accuracy, because more of the liability signal can be inferred from observed variants. We empirically verified that the advantage of LEAP over other methods increased with heritability, with noticeable power gains for diseases with heritability ≥25% (**Fig. 1** and **Supplementary Fig. 7**). We also performed a series of experiments to demonstrate that LEAP outperforms other methods under diverse levels of population structure, family relatedness, polygenicity and covariate effects (**Supplementary Figs. 8**-**15** and **Supplementary Note**).

To evaluate performance on real data, we analyzed nine disease data sets from the Wellcome Trust case control consortium (WTCCC)[5, 14, 15]. Measuring power for real data sets is an inherently difficult task because the identities of true causal single nucleotide polymorphisms (SNPs) are unknown. Evaluating type I error rates for real data is also a difficult task, because inflation of P values may stem from either sensitivity to confounding, or from high polygenicity of the studied trait[16].

As an approximate measure for type I error, we verified that the proportion of SNPs having $p<0.05$ and $p<10^{-5}$, and that are not within 2 Mbp of SNPs reported to be associated with the disease in previous studies, is comparable under LEAP and under a standard LMM. As an approximate measure for power, we computed normalized test statistics for SNPs that tag known risk variants from the US National Human Genome Research Institute catalog[17] as a 'bronze standard'.

LEAP demonstrated robustness to confounding, and was significantly more powerful than a standard LMM ($P<0.05$) in five out of the six rare phenotypes, with prevalence less than 1% (**Fig. 2** and **Supplementary Tables 2** and **3;** results for hypertension are omitted from Fig. 2 owing to the low number of known risk variants). As expected from the simulations, the advantage of LEAP increased with sample size and with confounding. Thus, only a small advantage was observed in the WTCCC1 data sets, which contain about 4,500 individuals per data set and little population structure or family relatedness, whereas a significantly greater advantage was observed in the larger and more confounded WTCCC2 data sets.

In the highly confounded multiple sclerosis (MS) data set[5], LEAP obtained a mean increase of more than 8% over an LMM in test statistics of tag SNPs, and an even greater advantage over other methods, while demonstrating robustness to confounding. All genome-wide significant loci identified by LEAP and LMM, having $p<5\times10^{-8}$, have previously been reported to be associated with MS in meta-analyses. In contrast, Linreg+PCs and Linreg identified 2 and 508 previously unidentified significant loci, respectively.

There are several avenues for future research. First, liabilities follow a truncated multivariate normal distribution, and thus their likelihood cannot be computed by an LMM without model misspecification, even if they are perfectly estimated. It may be possible to modify the objective function of LEAP so that its estimated liabilities follow a normal distribution[18]. Second, liability estimation may be improved by improving kinship estimation in ascertained studies (**Supplementary Note**). Finally, liability estimation may be improved by adopting richer models with a heterogeneous effect-size variance[19, 20].

## Acknowledgements
This work was supported by the Israeli Science Foundation. This study makes use of data generated by the Wellcome Trust Case Control Consortium. A full list of the investigators who contributed to the generation of the data is available from www.wtccc.org.uk. Funding for the project was provided by the Wellcome Trust under award 076113. The MS and UC data sets were filtered by Alexander Gusev. We thank N. Zaitlen and S. Rosset for helpful discussions.

**Conflict of interest**: C.L. and D.H. performed work on this manuscript while employed by Microsoft.

# Figures

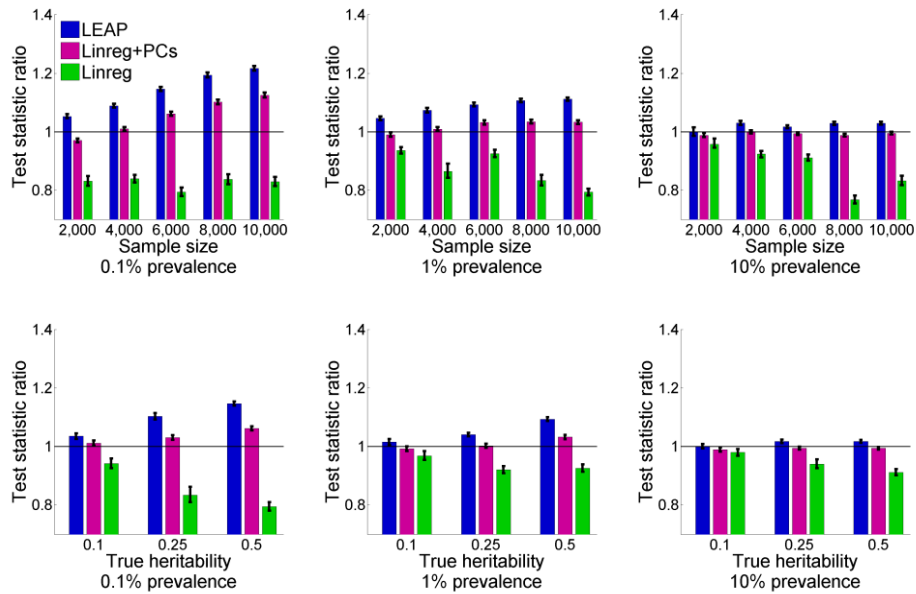

**Figure 1:** Synthetic data demonstrating that the power of LEAP increases with sample size and disease heritability. The values shown are the mean ratio of normalized test statistics for causal SNPs between each evaluated method and a standard LMM, under different sample sizes (top row) and heritability levels (bottom row), and 95% confidence intervals. Larger mean ratios indicate higher power. Values above the horizontal line indicate that a method has test statistics that are on average greater than that of a standard LMM.

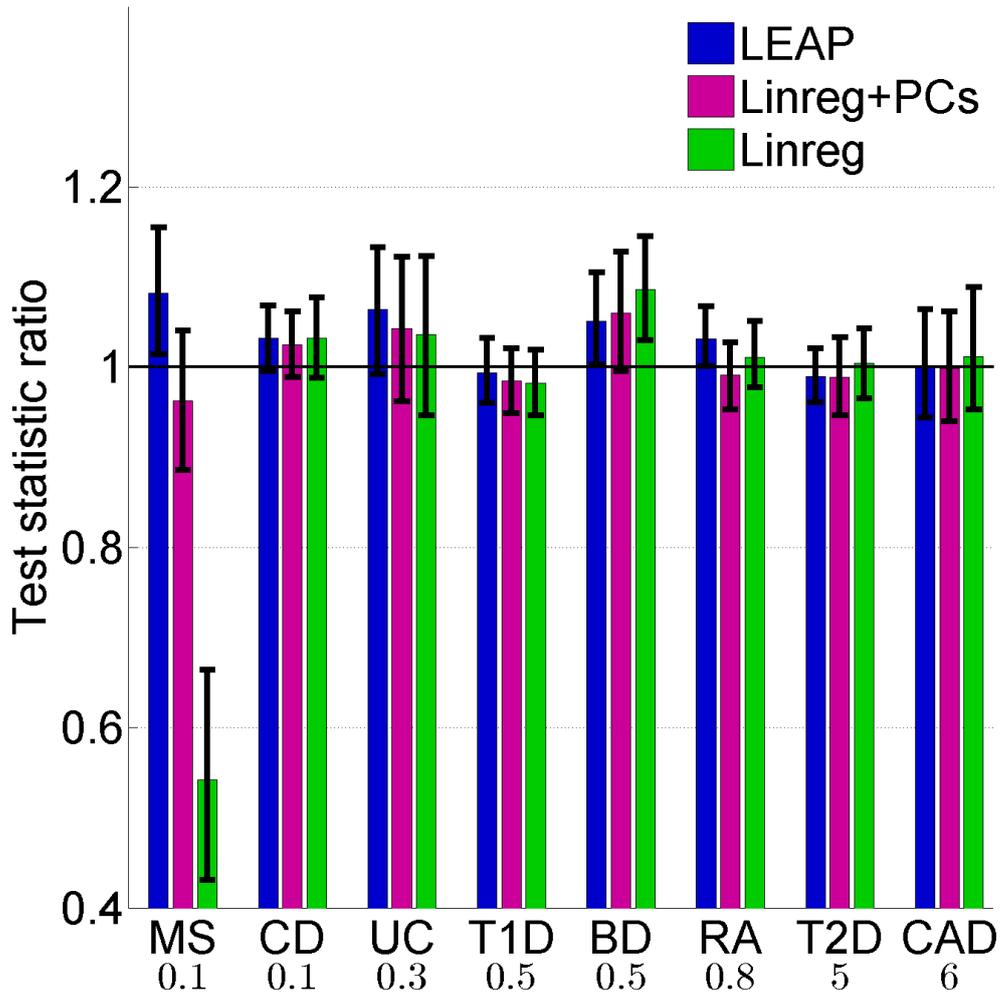

**Figure 2**: Analysis of real data sets with LEAP and other methods. The values shown are the mean ratio of normalized test statistics of SNPs tagging known variants between each evaluated method and a standard LMM, and its 95% confidence interval. A higher mean ratio indicates higher power. Values above the horizontal line indicate that a method has test statistics that are on average greater than that of a standard LMM. The number of tag SNPs is 44, 38, 20, 11, 19, 35, 17, and 15 for multiple sclerosis (MS), Crohn's disease (CD), ulcerative colitis (UC), type 1 diabetes (T1D), bipolar disorder (BD), rheumatoid arthritis (RA), type 2 diabetes (T2D), and coronary artery disease (CAD), respectively. The prevalence of each disease is shown below its name (in percent units).

# Online Methods

## LEAP Overview

The LEAP procedure is composed of four parts, which are now briefly overviewed, with detailed explanations following below.

1. Heritability estimation: The heritability of a trait quantifies the degree to which it is driven by genetic factors[2, 21, 22]. Several methods for heritability estimation in case-control studies have been proposed recently[2, 22]. We adopt the method of ref.[2], which directly models the ascertainment procedure.
2. Liability estimation: Using the heritability estimate, we fit a regularized Probit model, to estimate the effect size of each genetic variant on the liability. We use the Probit model to estimate liabilities for the sample individuals.
3. Association testing: The liability estimate is used as an observed phenotype in a GWAS context. Genetic variants are tested for association with this estimate via a standard LMM. The LMM is fitted using the heritability estimate, as described below.

We first provide a brief overview of the liability threshold model, and then derive a corresponding liability estimation model. Detailed derivations are found in the **Supplementary Note**.

## The Liability Threshold Model

According to the liability threshold model, every individual $i$ is associated with a latent normally distributed variable $l_i \sim N(0,1)$, such that cases are individuals with $l_i > t$, where t is the liability cutoff for a particular trait of interest. Assuming a trait with a population prevalence K, t is given by $\Phi^{-1}(1-K)$, where $\Phi^{-1}(\cdot)$ is the inverse cumulative probability density of the standard normal distribution. We decompose the liability $l_i$ into $l_i = g_i + e_i$, where $g_i$ and $e_i$ are the genetic and environmental components of the liability, respectively. Under standard assumptions, $e_i \sim N(0, \sigma_e^2)$ and $g_i = X_i^T \beta$, where $X_i$ is an $m \times 1$ vector of m standardized genetic variants carried by individual $i$, and $\beta$ is an $m \times 1$ vector of effect sizes, which follows the distribution $\beta \sim N\left(0, \frac{\sigma_g^2}{m} I\right)$, where I is the identity matrix. The genetic and environmental variances $\sigma_g^2$ and $\sigma_e^2$ are closely related to the narrow-sense heritability of a trait[2], defined as $\sigma_g^2 / (\sigma_g^2 + \sigma_e^2)$.

## Liability Estimation

As discussed in the **Supplementary Note**, testing for associations under the liability threshold model requires integrating the underlying liability vector over its support. Motivated by this observation, we propose approximating such association testing by selecting a liability estimator and treating it as the observed phenotype vector. A good liability estimator has values close to the true, unobserved, underlying liability. Thus, the problem is equivalent to inferring the value of an unknown continuous variable with a known distribution.

We estimate liabilities via a maximum a posteriori (MAP) estimator, which estimates the MAP of the effect sizes of all genetic variants conditional on the phenotypes, genotypes and the disease prevalence. The likelihood to maximize can be written as

$$P\left(\beta; 0, \frac{\sigma_g^2}{m}I\right) \prod_{i \in controls} \Phi(t - X_i^T\beta; 0, \sigma_e^2) \prod_{i \in cases} \left(1 - \Phi(t - X_i^T\beta; 0, \sigma_e^2)\right). \quad (1)$$

Taking the logarithm and using the normal distribution definition, the quantity to maximize is

$$\sum_{i \in controls} \log \Phi(t - X_i^T\beta; 0, \sigma_e^2) + \sum_{i \in cases} \log \left(1 - \Phi(t - X_i^T\beta; 0, \sigma_e^2)\right) - \frac{1}{2\sigma_g^2/m} \sum_j \beta_j^2 + W \quad (2)$$

where $W$ is a quantity that does not depend on $\beta$ and can thus be ignored. This problem is equivalent to Probit regression[23] with L2 regularization and a pre-specified offset term, and can thus be solved using standard techniques (**Supplementary Note**). Unlike typical uses of such models, here the regularization parameter is known in advance, given a value for $\sigma_g^2$.

The MAP for $\hat{g}_i$ is given by $\hat{g}_i = X_i^T \hat{\beta}$, where $\hat{\beta}$ is the MAP of $\beta$. Given the MAP $\hat{g}_i$, $\hat{l}_i$ is equal to $\hat{g}_i$ if individual i is a control and $\hat{g}_i < t$ or if individual i is a case and $\hat{g}_i > t$, and is equal to t otherwise. This follows because $e_i$ has a zero-mean normal distribution.

### *Dimensionality Reduction*

A straightforward solution of the optimization problem presented above is difficult owing to its high dimensionality, which is equal to the number of genotyped variants. Fortunately, the problem can be reformulated as a lower dimensional problem, with dimensionality equal to the number of individuals.

The equivalence stems from the fact that the genotypes matrix $X$ can be represented in terms of the eigenvectors of its covariance matrix alone. To see this, we rewrite

Expression 2 as follows
$$f(X\beta) - \frac{1}{2\sigma_g^2/m}\beta^T\beta \qquad (3)$$

where $f(X\beta)$ is a function that depends on $\beta$ only through the product $X\beta$. Consider the singular value decomposition (SVD) of $X$, given by
$$X = USV^T \qquad (4)$$

where $U$ is the matrix of the eigenvectors of $XX^T$, and $V$ is orthonormal. Denote $Z = US$ and $\beta_Z = V^T\beta$. Owing to the orthonormality of $V$, the following equations hold:

$$\beta_Z^T\beta_Z = \beta^T\beta. \qquad (5)$$

$$Z\beta_Z = X\beta. \qquad (6)$$

Therefore, Expression 3 can be rewritten as
$$f(Z\beta_Z) - \frac{1}{2\sigma_g^2/m}\beta_Z^T\beta_Z. \qquad (7)$$

Denoting the number of individuals and genotyped variants by $n$ and $m$, respectively, and assuming $m > n$ and that the columns of $Z$ are ordered according to the magnitude of their respective eigenvalues, then all columns of $Z$ except for the leftmost $n$ ones are equal to zero. Consequently, the vector $Z\beta_Z$ depends only on the top $n$ entries of the vector $\beta_Z$, and thus all the other entries can be set to 0.

We conclude that the quantity in equation (2) can be maximized by considering only the non-zero components of the matrix $Z$ and the vector $\beta_Z$, which have dimensionalities $n \times n$ and $n$, respectively. In contrast, the original formulation of the problem uses the matrix $X$ and the vector $\beta$, which have dimensionalities $n \times m$ and $m$, respectively. The original effect sizes are given by $\beta = V\beta_Z$. However, they are not needed in practice, since the liabilities estimator can be computed using $\beta_Z$ directly.

Finally, we note that when performing GWAS, the matrix $Z$ is typically computed regardless of whether LEAP is employed, and is thus available at no further computational cost. This results from the close relation between the SVD of $X$ and the eigendecomposition of the matrix $XX^T$. Namely, given the eigendecomposition $XX^T = US^2U^T$, the matrix $Z$ is given by $Z = US$, where $S$ is the matrix of the componentwise square roots of the entries of $S^2$. In GWAS, the eigendecomposition of $XX^T$ is computed both when using an LMM[24] and when performing regression using principal component covariates[12], and is thus available for use in LEAP at no further computational cost.

*Use in GWAS*

LEAP uses liability estimates by treating them as observed continuous phenotypes in an LMM. Three difficulties that must be dealt with are accurate fitting of the LMM parameters, avoiding testing SNPs for association with the liability estimator that they helped estimate, and dealing with family relatedness. We now describe solutions to these difficulties.

The difficulty of parameter estimation stems from the non-normality of the liability under case-control sampling. This non-normality arises because in rare diseases, the majority of cases share a similar liability close to the cutoff. Parameter estimation can be suboptimal in such settings. The most important parameter that is fitted in LMMs is the variances ratio $\delta = \sigma_e^2/\sigma_g^2$. Given this parameter, all other parameters can be evaluated via closed form formulas[24]. There is a close connection between this parameter and the narrow-sense heritability, $h^2 = \sigma_g^2/(\sigma_g^2 + \sigma_e^2)$, expressed via $\delta = 1/h^2 - 1$. We therefore fit this parameter by estimating the heritability using the method of ref.[2], as described in the **Supplementary Note**.

A second difficulty arises because SNPs should not be tested for association with a liability estimator that they helped estimate. Otherwise the test statistic for these SNPs will be inflated, because they can always account for some of the liability variance. Similarly, SNPs in linkage disequilibrium with a tested SNP should also not participate in the liability estimation. To prevent such inflation, we estimate liabilities on a per-chromosome basis. For every chromosome, the liability is estimated using all SNPs except for the ones on the chromosome. The SNPs on the excluded chromosome are then tested for association using this liability estimator. We note that LMM-based GWAS typically compute the eigendecomposition of the covariance matrix on a per-chromosome basis as well, in order to prevent a SNP from incorrectly affecting the null likelihood (the phenomenon termed *proximal contamination*[7, 24, 25]). LEAP can make use of these available eigendeompositions for dimensionality reduction - thus incurring no computational cost other than the liability estimation procedure itself.

A third difficulty arises when the data is confounded by family relatedness. The presence of related individuals can lead to biased effect size estimates, and consequently to a biased liability estimator. We deal with this difficulty by excluding related individuals from the parameter estimation stage of the MAP computation. We employ a greedy algorithm, where at each stage we exclude the individual having the largest number of related individuals with correlation coefficient >0.05. After fitting the model, we estimate liabilities for the excluded individuals as well. We note that population structure does not present similar problems, because it is naturally captured by top principal components[3, 12, 26], which are fitted in the MAP computation.

*Data Simulation*

All experiments reported in this paper are based on a uniform data generation procedure that can simulate different settings via a variety of parameters. In these simulations, each individual carried 60,100 SNPs that do not affect the phenotype, as well as 50-5,000 causal SNPs with normally distributed effect sizes. Population structure was simulated via the Balding-Nichols model[27], which generates populations with genetic divergence measured via Wright's $F_{ST}$[13]. Family relatedness was simulated by generating various numbers of sib-pairs in one of the two populations, as in ref.[3]. To simulate ascertainment, we generated $3,000/K$ individuals and a latent liability value for every individual, where $K$ is the disease prevalence. We then determined the $1-K$ percentile of the liabilities, and generated new individuals until 50% of the sample had liabilities exceeding this cutoff[9]. Unless otherwise noted, all simulations use 6,000 individuals, $F_{ST}=0.01$ and 30% of the individuals in one of the two populations are sibling pairs. In all experiments, ten data sets were generated for each unique combination of settings. A detailed description of the simulation procedure and its default parameters is provided in the **Supplementary Note**.

## Software and Code Availability

LEAP is available to download from https://github.com/omerwe/LEAP.
LEAP has the same memory requirements as the FaST-LMM package[20], and is computationally efficient. On a 2 GHz CPU, it can accurately estimate liabilities for samples as large as 50,000 individuals in fewer than 5 min.

# Supplementary Tables

**Supplementary Table S1.** Type I error rates for synthetic experiments. The table shows the type I error rates of the tested methods on synthetic data sets with 6,000 individuals and 0.1% prevalence, using various $F_{ST}$ levels and proportions of individuals in one of the two populations who are part of a sib-pair (denoted as S). For each method we report its genomic control inflation factor[28] $\lambda_{GC}$, defined as the ratio of expected to observed median test statistic in $\chi^2$ space, and the average ratio of the actual type I error rates at p=0.05 and p=$10^{-5}$ over their expected values. For every measure we also report its 95% confidence interval (CI), obtained via 1,000 bootstrap samples over the test statistics of every data set. Each result is averaged across 10 data sets.

|  |  | $F_{ST}$=0.01 | | | S=30% | |
|---|---|---|---|---|---|---|
|  |  | S=0% | S=3% | S=30% | $F_{ST}$=0 | $F_{ST}$=0.05 |
| LEAP | $\lambda_{GC}$ | 0.95 | 0.96 | 0.99 | 0.99 | 1.03 |
|  | $\lambda_{GC}$ CI | 0.94-0.97 | 0.94-0.97 | 0.98-1.01 | 0.98-1.01 | 1.02-1.05 |
|  | p=0.05 | 0.858 | 0.875 | 0.950 | 0.948 | 0.982 |
|  | p=0.05 CI | 0.830-0.885 | 0.848-0.902 | 0.922-0.979 | 0.920-0.976 | 0.954-1.012 |
|  | p=$10^{-5}$ | 0.499 | 0.832 | 0.832 | 0.666 | 0.666 |
|  | p=$10^{-5}$ CI | 0.000-1.498 | 0.000-2.496 | 0.000-2.496 | 0.000-1.997 | 0.000-1.997 |
| LMM | $\lambda_{GC}$ | 1.00 | 1.00 | 1.03 | 1.03 | 1.03 |
|  | $\lambda_{GC}$ CI | 0.98-1.01 | 0.98-1.01 | 1.01-1.05 | 1.01-1.05 | 1.02-1.05 |
|  | p=0.05 | 0.962 | 0.972 | 1.040 | 1.016 | 1.041 |
|  | p=0.05 CI | 0.933-0.991 | 0.944-1.001 | 1.010-1.070 | 0.987-1.046 | 1.011-1.071 |
|  | p=$10^{-5}$ | 0.666 | 0.666 | 1.331 | 0.832 | 0.499 |
|  | p=$10^{-5}$ CI | 0.000-1.830 | 0.000-1.997 | 0.000-3.827 | 0.000-2.329 | 0.000-1.498 |
| Linreg +PCs | $\lambda_{GC}$ | 1.01 | 1.02 | 1.06 | 1.06 | 1.07 |
|  | $\lambda_{GC}$ CI | 1.00-1.03 | 1.01-1.04 | 1.05-1.08 | 1.04-1.07 | 1.05-1.08 |
|  | p=0.05 | 1.007 | 1.017 | 1.111 | 1.099 | 1.122 |
|  | p=0.05 CI | 0.978-1.036 | 0.988-1.047 | 1.080-1.142 | 1.069-1.130 | 1.090-1.152 |
|  | p=$10^{-5}$ | 1.331 | 1.165 | 1.165 | 1.664 | 1.331 |
|  | p=$10^{-5}$ CI | 0.000-3.170 | 0.158-2.995 | 0.000-3.161 | 0.000-4.326 | 0.166-3.328 |
| Linreg | $\lambda_{GC}$ | 1.56 | 2.12 | 1.79 | 1.06 | 7.77 |
|  | $\lambda_{GC}$ CI | 1.54-1.59 | 2.09-2.15 | 1.76-1.81 | 1.04-1.07 | 7.64-7.89 |
|  | p=0.05 | 2.220 | 3.220 | 2.671 | 1.098 | 6.042 |
|  | p=0.05 CI | 2.179-2.260 | 3.175-3.266 | 2.627-2.713 | 1.067-1.129 | 5.990-6.094 |
|  | p=$10^{-5}$ | 192.3 | 759.7 | 359.7 | 1.331 | 9314.0 |
|  | p=$10^{-5}$ CI | 170.0-214.5 | 719.0-802.2 | 329.9-389.2 | 0.000-3.328 | 9207-9424 |

**Supplementary Table S2.** Summary statistics and experimental results for the multiple sclerosis (MS) and ulcerative colitis (UC) data sets. For each data set, we considered two lists of tag SNPs: the list of tag SNPs reported in ref.[7], and the subset of those SNPs with p<0.01 in at least one of the methods. For each method we report (a) the ratio between the mean of test statistics of tag SNPs and the corresponding mean of a standard LMM and its 95% confidence interval (CI), (b) the mean of the ratios of test statistics of tag SNPs with those of a standard LMM, using the subset of tag SNPs, and its 95% CI (c) the P value of the ratio of the means, (d) the inflation factor $\lambda_{GC}$, and (e) the ratio of actual type I error rates at p=0.05 and p=$10^{-5}$ over their expected values. The type I error rate is computed under the (probably incorrect) assumption that a SNP is unassociated with the disease if it is at least 2M base pairs away from every previously reported associated SNP.

|  |  | MS | UC |
|---|---|---|---|
| **Prevalence** |  | 0.1% | 0.3% |
| **#cases** |  | 10204 | 2697 |
| **#controls** |  | 5429 | 5652 |
| **#tag SNPs** |  | 75 | 25 |
| **#filtered tag SNPs** |  | 44 | 20 |
| LEAP | Ratio of Means | 1.04 | 1.07 |
|  | 95% CI | 1.00-1.09 | 1.00-1.14 |
|  | Mean of Ratios | 1.08 | 1.06 |
|  | 95% CI | 1.01-1.16 | 0.99-1.13 |
|  | P value | 0.035 | 0.027 |
|  | $\lambda_{GC}$ | 1.18 | 1.08 |
|  | p=0.05 | 1.46 | 1.17 |
|  | p=$10^{-5}$ | 13.96 | 5.64 |
| LMM | $\lambda_{GC}$ | 1.20 | 1.14 |
|  | p=0.05 | 1.48 | 1.33 |
|  | p=$10^{-5}$ | 12.72 | 7.67 |
| Linreg+PCs | Ratio of Means | 0.91 | 1.05 |
|  | 95% CI | 0.84-0.98 | 0.98-1.12 |
|  | Mean of Ratios | 0.96 | 1.04 |
|  | 95% CI | 0.89-1.04 | 0.96-1.12 |
|  | P value | 0.99 | 0.10 |
|  | $\lambda_{GC}$ | 1.26 | 1.10 |
|  | p=0.05 | 1.60 | 1.22 |
|  | p=$10^{-5}$ | 39.41 | 6.32 |
| Linreg | Ratio of Means | 0.51 | 1.03 |
|  | 95% CI | 0.44-0.59 | 0.95-1.12 |
|  | Mean of Ratios | 0.54 | 1.04 |
|  | 95% CI | 0.43-0.66 | 0.95-1.12 |
|  | P value | 1.00 | 0.21 |
|  | $\lambda_{GC}$ | 3.86 | 1.16 |
|  | p=0.05 | 6.36 | 1.35 |
|  | p=$10^{-5}$ | 2670.3 | 7.00 |

**Supplementary Table S3.** Summary statistics and experimental results for the Wellcome Trust 1 data sets. Notations are the same as for Table S2. The phenotypes are Crohn's disease (CD), rheumatoid arthritis (RA), type 1 diabetes (T1D), bipolar disorder (BD), type 2 diabetes (T2D), coronary artery disease (CAD) and hypertension (HT). Results for mean of ratios for HT are not shown because its analysis includes only 3 SNPs, yielding an unreliable estimate of this quantity.

|  |  | CD | RA | T1D | BD | T2D | CAD | HT |
|---|---|---|---|---|---|---|---|---|
| **Prevalence** |  | 0.1% | 0.8% | 0.5% | 0.5% | 5% | 6% | 30% |
| **#cases** |  | 1726 | 1834 | 1945 | 1850 | 1902 | 1905 | 1926 |
| **#controls** |  | 2925 | 2925 | 2925 | 2925 | 2925 | 2925 | 2925 |
| **#tag SNPs** |  | 63 | 31 | 34 | 98 | 41 | 50 | 20 |
| **#filtered tag SNPs** |  | 38 | 11 | 19 | 35 | 17 | 15 | 3 |
| LEAP | Ratio of Means | 1.03 | 1.04 | 0.99 | 1.03 | 0.98 | 1.00 | 0.95 |
|  | 95% CI | 1.00-1.06 | 1.02-1.07 | 0.97-1.02 | 1.00-1.07 | 0.95-1.01 | 0.96-1.04 | 0.86-1.0 |
|  | Mean of Ratios | 1.03 | 1.03 | 0.99 | 1.05 | 0.99 | 1.00 | - |
|  | 95% CI | 1.00-1.07 | 1.00-1.07 | 0.96-1.03 | 1.00-1.11 | 0.96-1.02 | 0.94-1.06 | - |
|  | P value | 0.017 | 0.002 | 0.725 | 0.034 | 0.924 | 0.555 | 0.96 |
|  | $\lambda_{GC}$ | 1.07 | 1.02 | 1.02 | 1.05 | 1.02 | 1.02 | 0.98 |
|  | p=0.05 | 1.10 | 1.07 | 1.05 | 1.13 | 1.06 | 1.02 | 0.98 |
|  | p=$10^{-5}$ | 0.41 | 0.39 | 0.40 | 0.00 | 0.00 | 0.00 | 0.40 |
| LMM | $\lambda_{GC}$ | 1.08 | 1.03 | 1.04 | 1.09 | 1.07 | 1.08 | 1.06 |
|  | p=0.05 | 1.13 | 1.09 | 1.09 | 1.22 | 1.15 | 1.14 | 1.15 |
|  | p=$10^{-5}$ | 0.82 | 0.00 | 0.40 | 0.47 | 0.83 | 0.00 | 1.60 |
| Linreg+PCs | Ratio of Means | 1.03 | 1.00 | 0.98 | 1.02 | 0.96 | 1.00 | 1.04 |
|  | 95% CI | 0.99-1.06 | 0.97-1.04 | 0.96-1.01 | 0.98-1.07 | 0.93-1.01 | 0.96-1.04 | 0.97-1.16 |
|  | Mean of Ratios | 1.03 | 0.99 | 0.99 | 1.06 | 0.99 | 1.00 | - |
|  | 95% CI | 0.99-1.06 | 0.95-1.03 | 0.95-1.02 | 1.00-1.13 | 0.95-1.03 | 0.94-1.06 | - |
|  | P value | 0.075 | 0.431 | 0.928 | 0.189 | 0.946 | 0.514 | 0.166 |
|  | $\lambda_{GC}$ | 1.07 | 1.04 | 1.04 | 1.09 | 1.07 | 1.07 | 1.07 |
|  | p=0.05 | 1.11 | 1.10 | 1.10 | 1.23 | 1.16 | 1.13 | 1.18 |
|  | p=$10^{-5}$ | 0.41 | 0.78 | 0.00 | 0.00 | 0.42 | 2.04 | 2.00 |
| Linreg | Ratio of Means | 1.03 | 1.03 | 0.98 | 1.06 | 0.99 | 1.01 | 1.02 |
|  | 95% CI | 1.00-1.07 | 1.00-1.06 | 0.96-1.01 | 1.02-1.11 | 0.96-1.03 | 0.98-1.05 | 0.94-1.15 |
|  | Mean of Ratios | 1.03 | 1.01 | 0.98 | 1.09 | 1.00 | 1.01 | - |
|  | 95% CI | 0.99-1.08 | 0.98-1.05 | 0.95-1.02 | 1.03-1.15 | 0.97-1.04 | 0.95-1.09 | - |
|  | P value | 0.055 | 0.031 | 0.915 | 0.003 | 0.662 | 0.272 | 0.357 |
|  | $\lambda_{GC}$ | 1.11 | 1.03 | 1.04 | 1.12 | 1.08 | 1.08 | 1.07 |
|  | p=0.05 | 1.19 | 1.11 | 1.11 | 1.27 | 1.18 | 1.15 | 1.17 |
|  | p=$10^{-5}$ | 0.41 | 0.39 | 0.00 | 0.00 | 1.66 | 2.04 | 0.40 |

# Supplementary Figures

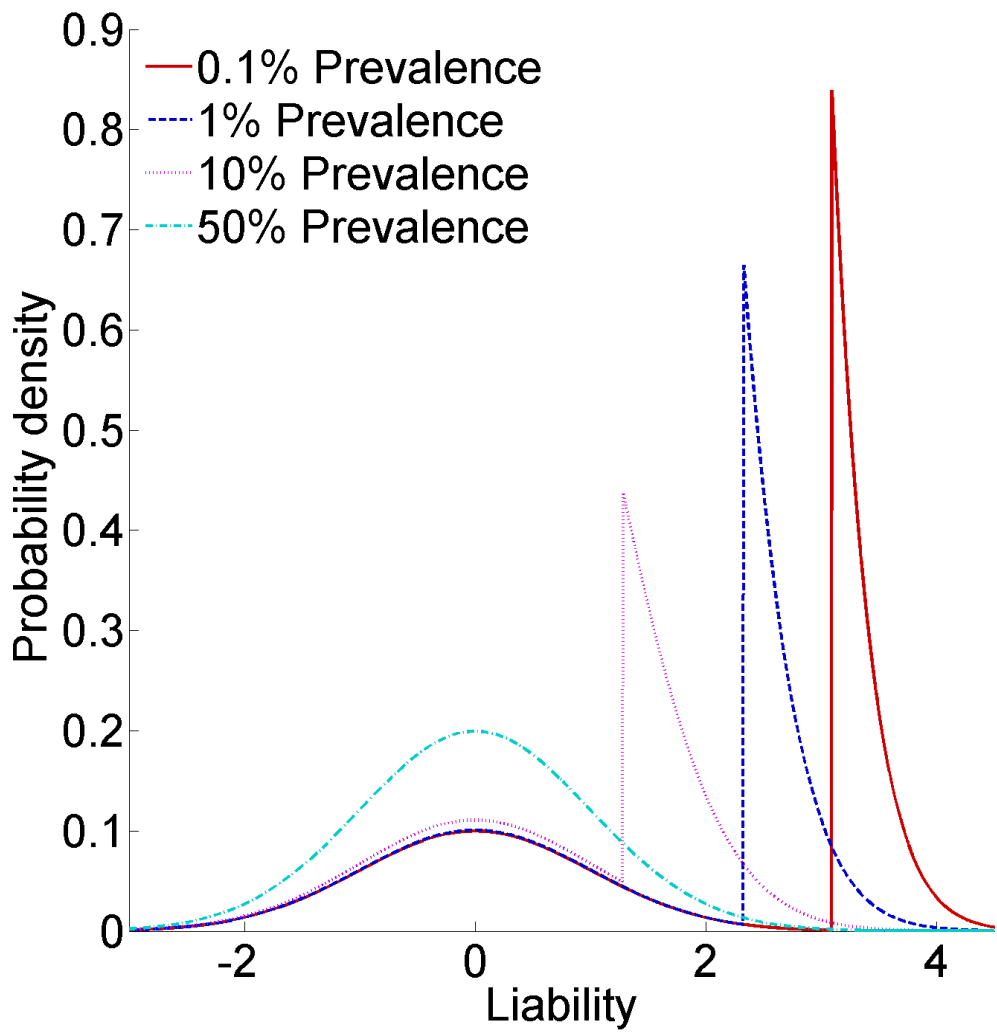

**Supplementary Figure 1: Liability distributions in balanced case-control data sets.** Individuals with liability greater than the prevalence-specific cutoff are cases, and the remainder are controls. The liabilities of controls and of cases follow a zero-mean normal distribution, conditioned on being smaller or greater than the liability cutoff, respectively. The distribution of case liabilities becomes increasingly sharply peaked as prevalence decreases.

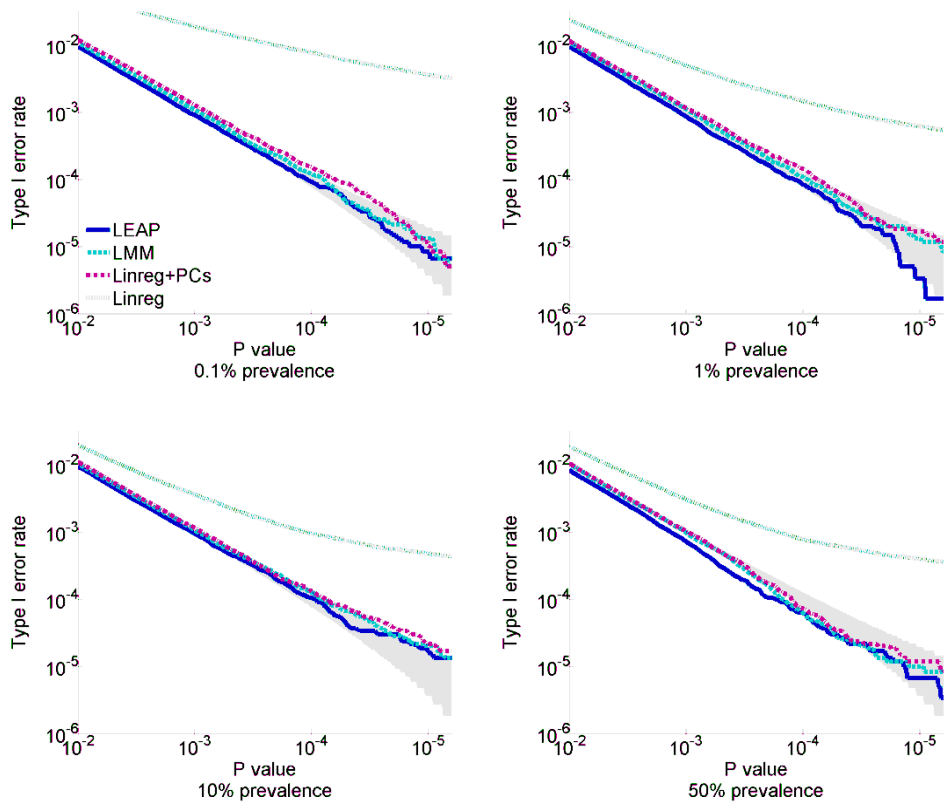

**Supplementary Figure 2: Type 1 error rates under different prevalence levels.** All experiments were run with FST=0.01 and samples where 30% of the individuals in one of the two populations are sib-pairs. The gray shaded area is the 95% confidence interval of the null distribution.

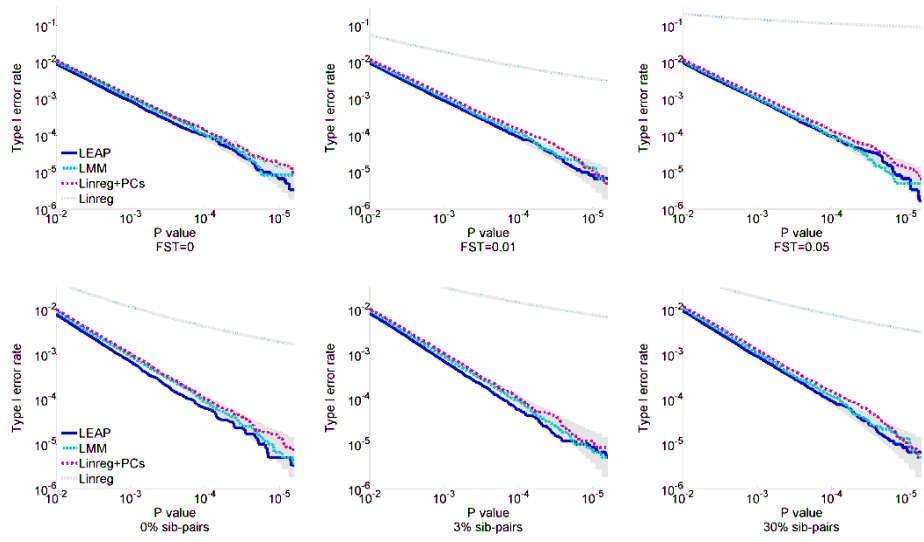

**Supplementary Figure 3:** Type 1 error rates under different population structure and family relatedness levels.

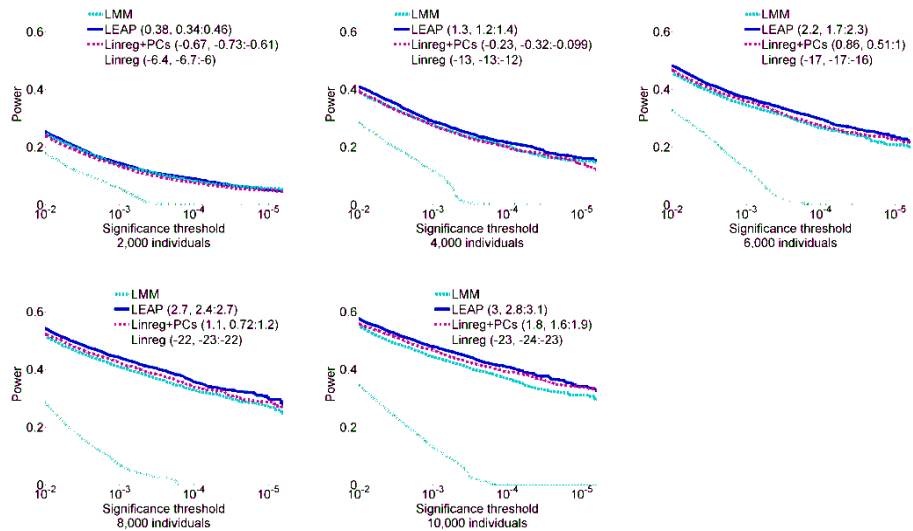

**Supplementary Figure 4: Power evaluations with different sample sizes, under 0.1% prevalence.** The mean relative increase in power of every method over an LMM is shown next to its name, in percentage units. For example, the number 3 indicates that a method has average power 3% greater than that of an LMM. Also shown is the 95% confidence interval of the mean increase, obtained via 10,000 bootstrap samples.

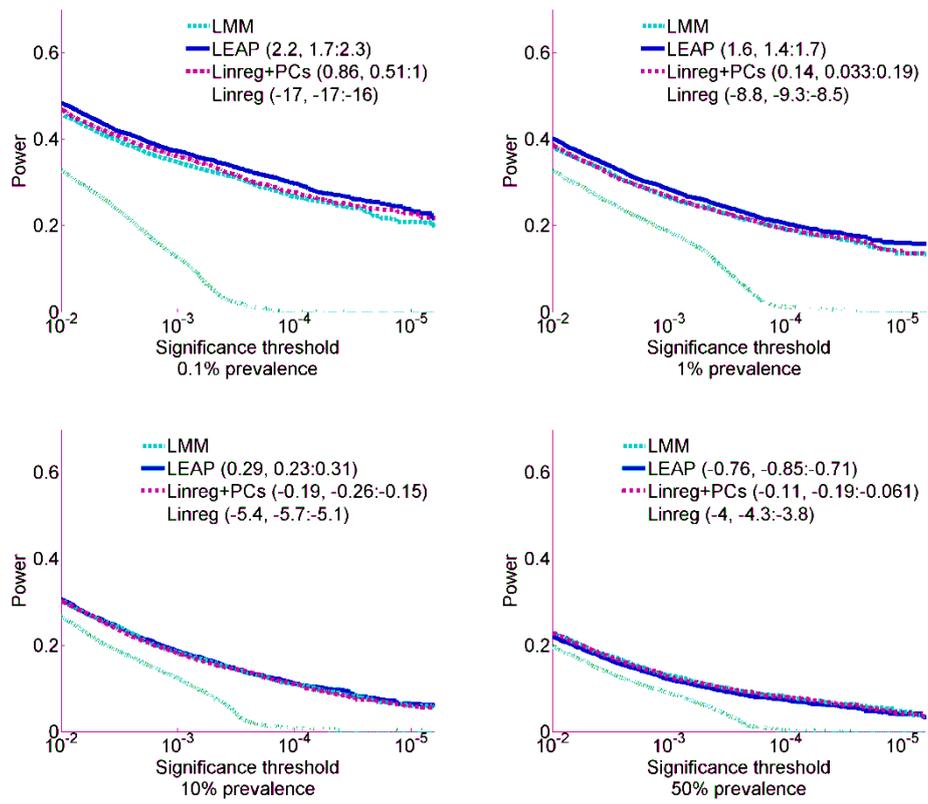

**Supplementary Figure 5: Power evaluations under different prevalence levels, with samples of 6,000 individuals.**

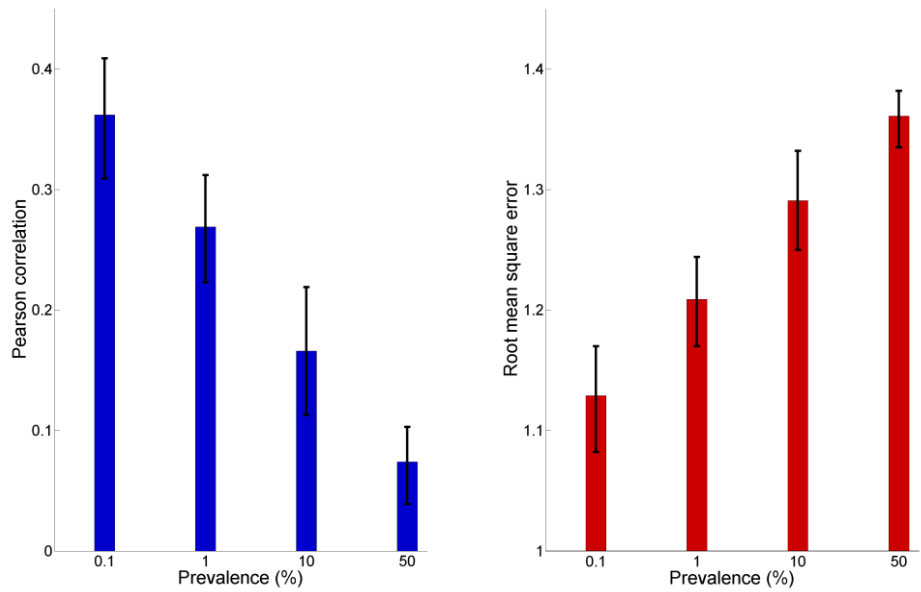

**Supplementary Figure 6: Similarity between the estimated and true liabilities of controls.** The figure shows results for data sets with 6,000 individuals, and their 95% confidence intervals (computed via 10-fold cross validation for each data set, averaged over 10 data sets). The similarity measures shown are the Pearson correlation and the root mean square error, after normalizing the liabilities to have zero mean and unit variance. The evaluation was applied only for controls, because liabilities of cases are trivial to estimate, as they are tightly clustered near the liability cutoff (Fig. S1).

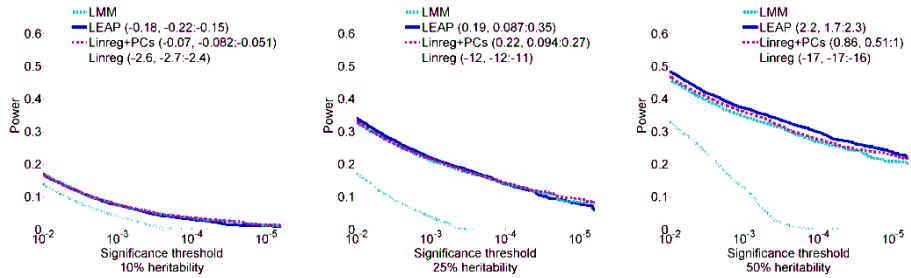

**Supplementary Figure 7: Power evaluations under different heritability levels.**

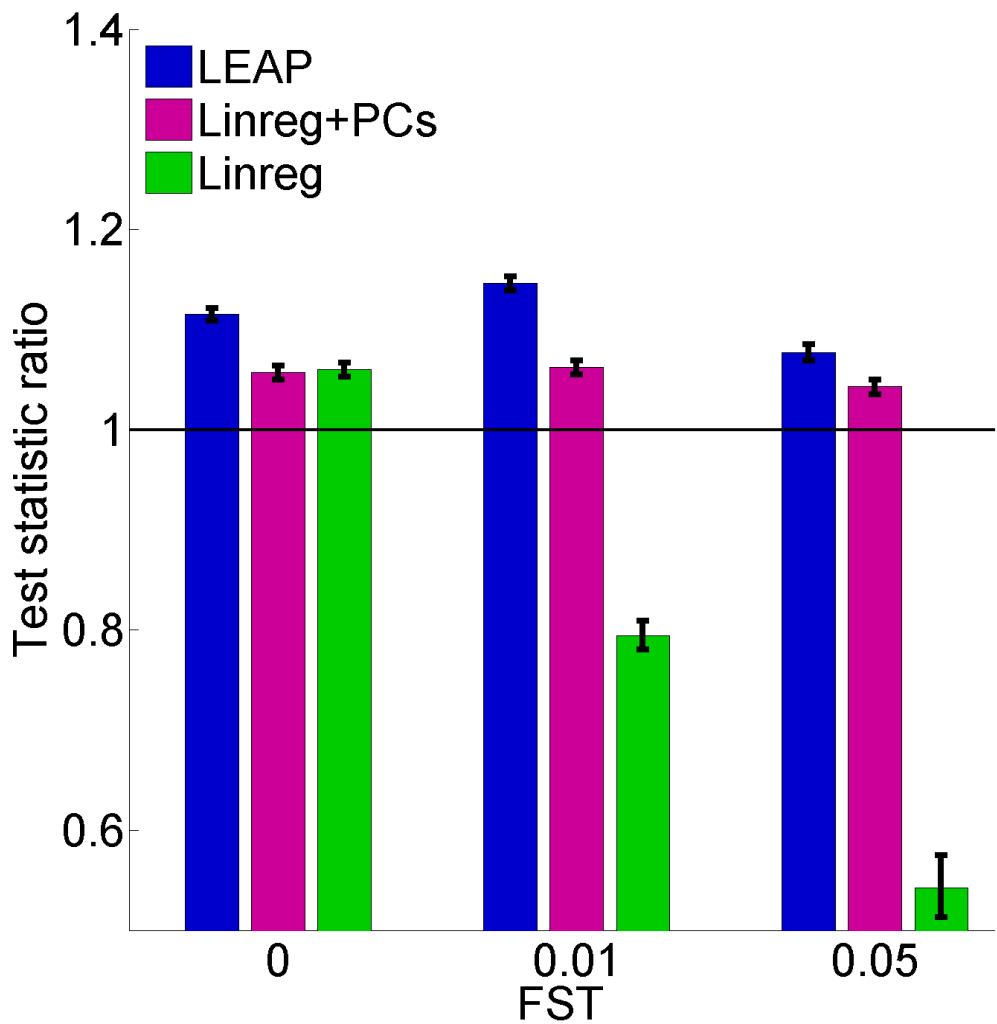

**Supplementary Figure 8: Population structure experiments.** The figure shows the mean ratio of normalized test statistics for causal SNPs between each evaluated method and an LMM under 0.1% prevalence, 30% sib-pairs, and various population structure levels.

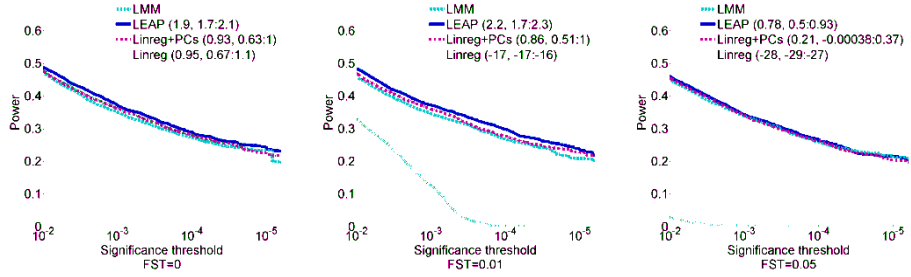

**Supplementary Figure 9: Power evaluations under different population structure levels.**

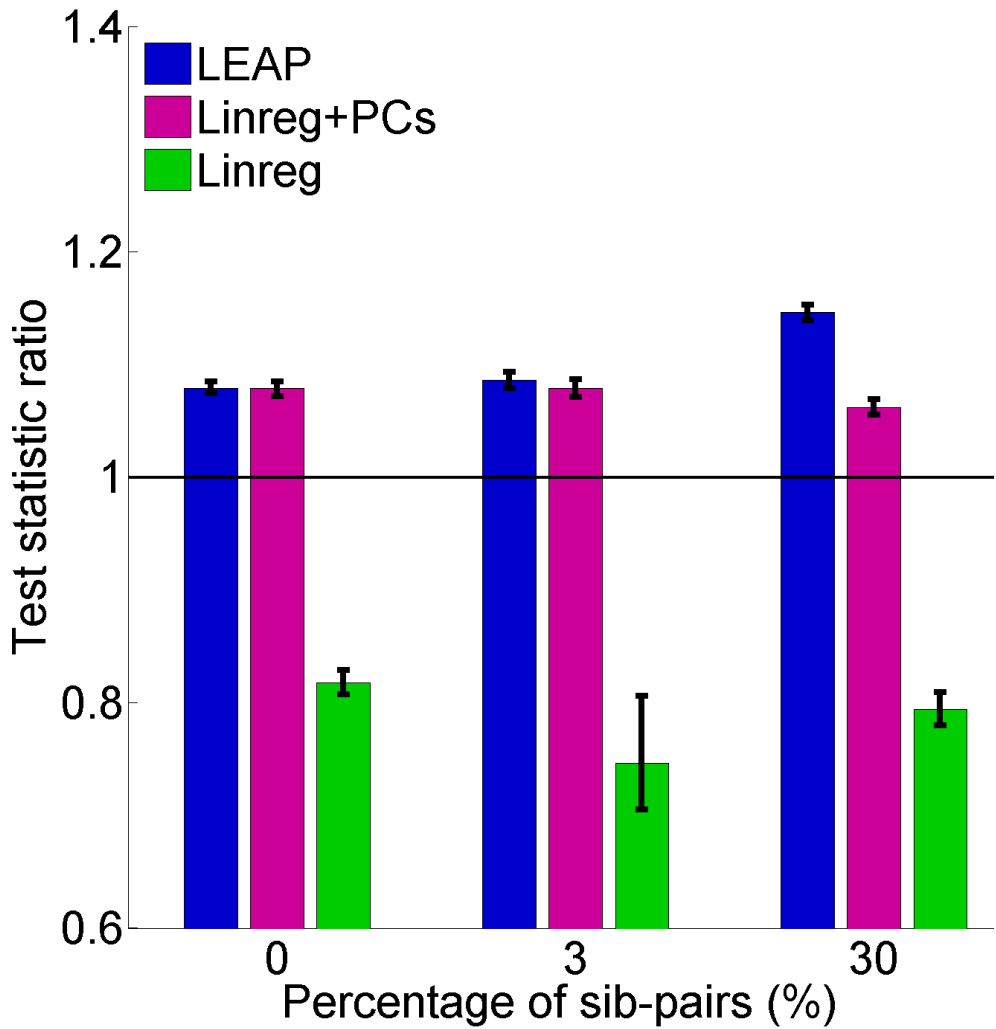

**Supplementary Figure 10: Family relatedness experiments.** The figure shows the mean ratio of normalized test statistics for causal SNPs between each evaluated method and an LMM under 0.1% prevalence, FST=0.01 and various percentages of individuals in one of the two populations who are sib-pairs.

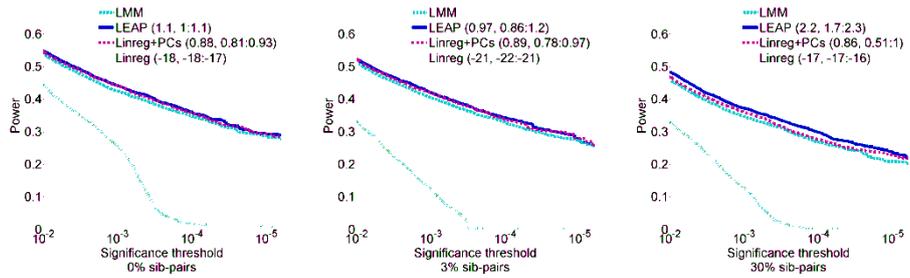

**Supplementary Figure 11: Power evaluations under different family relatedness levels.**

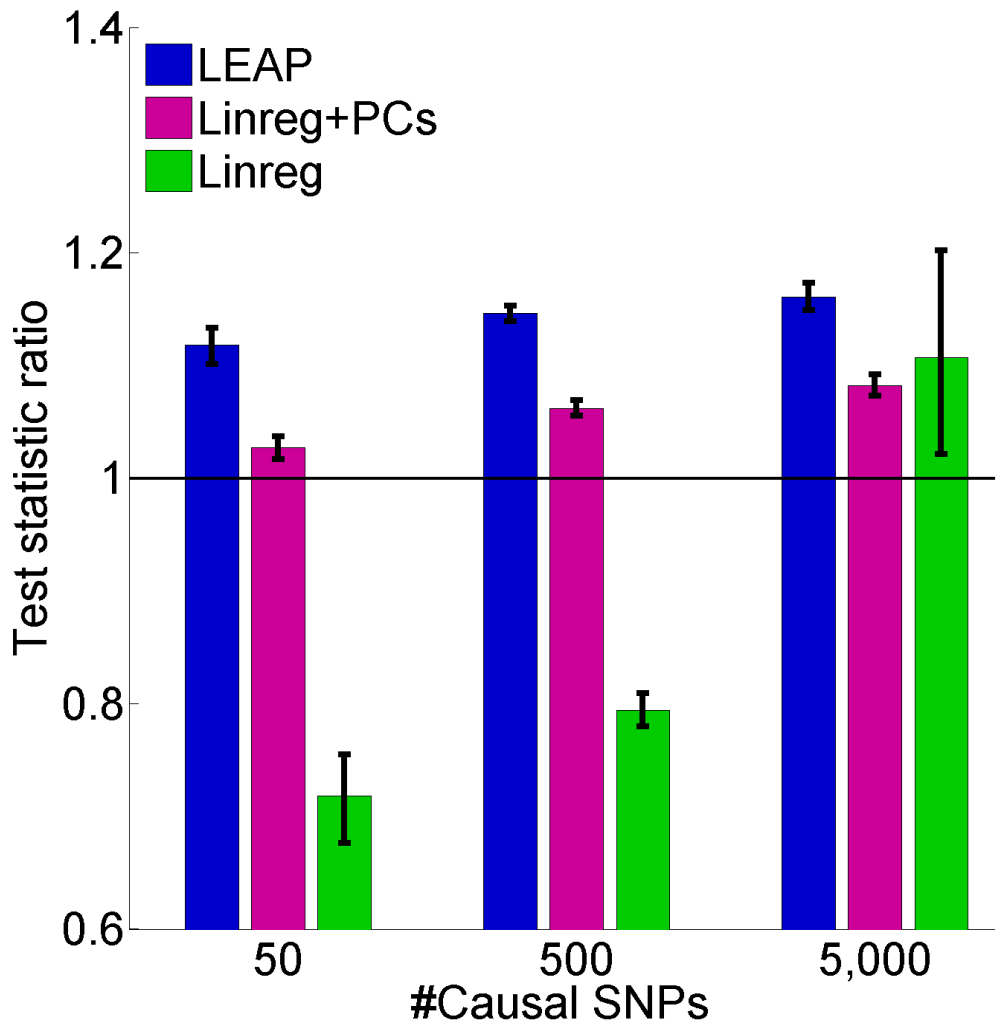

**Supplementary Figure 12: Polygenicity experiments.** The figure shows the mean ratio of normalized test statistics for causal SNPs between each evaluated method and an LMM under 0.1% prevalence and various numbers of causal SNPs.

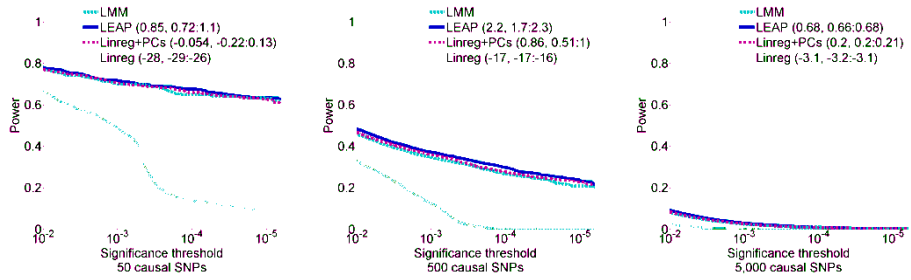

**Supplementary Figure 13: Power evaluations under different numbers of causal SNPs.**

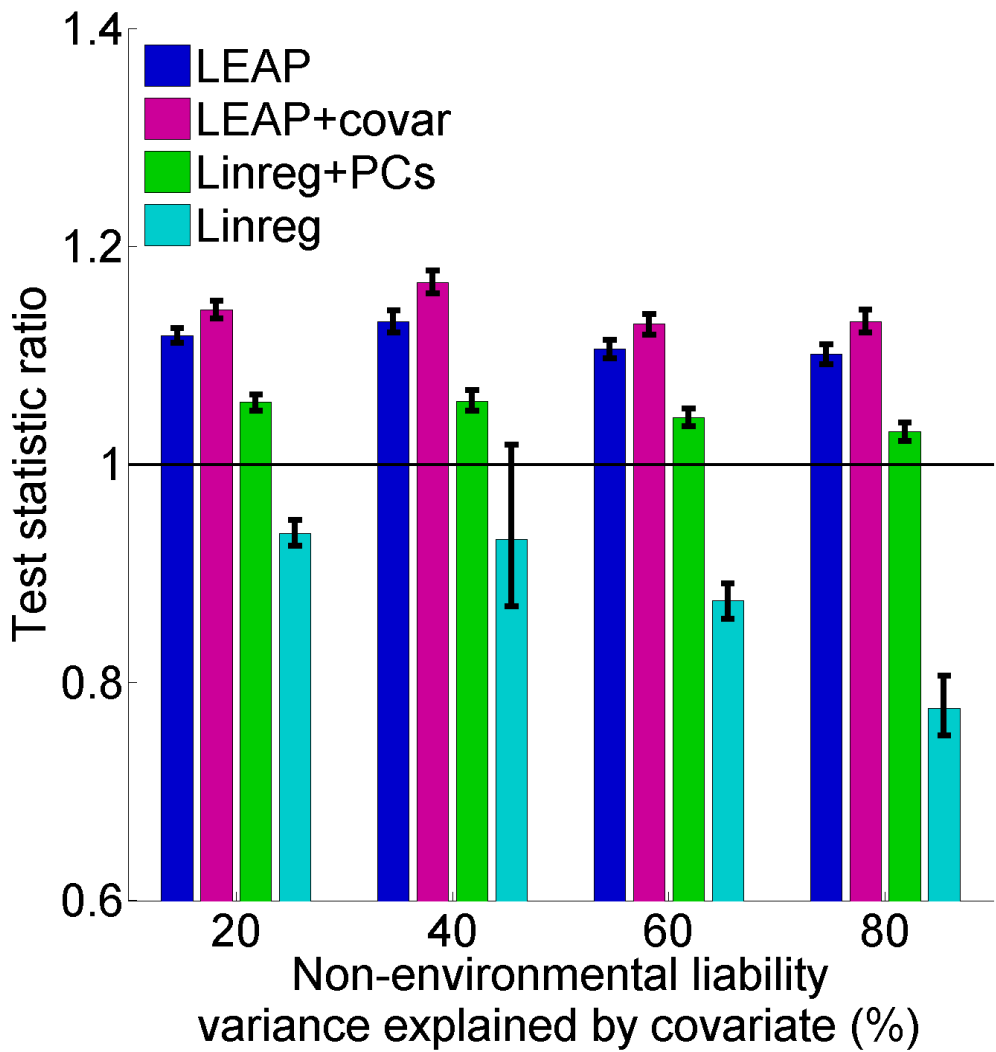

**Supplementary Figure 14: The mean ratio of normalized test statistics for causal SNPs between each evaluated method and an LMM, in the presence of covariates.** LEAP+covar is a variant of LEAP that uses covariates as well as genetic variants for liability estimation (Supplementary Note 2).

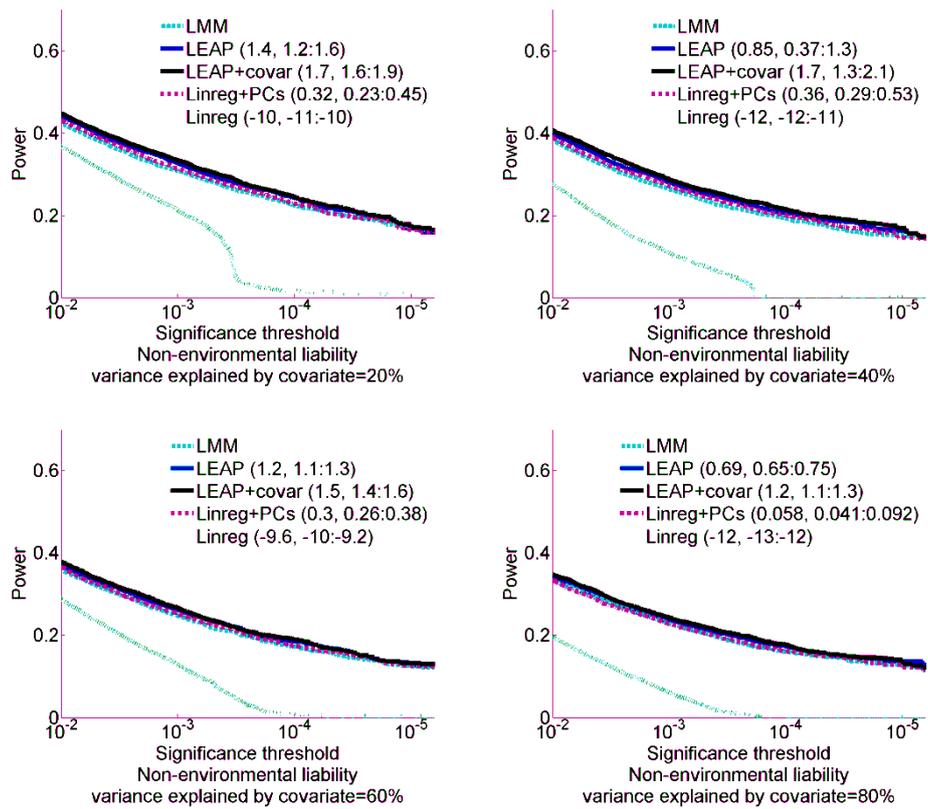

**Supplementary Figure 15: Power evaluations in the presence of covariates.**
LEAP+covar is a variant of LEAP that uses covariates as well as genetic variants for liability estimation (Supplementary Note 2).

# Supplementary Note 1 - Appendix

## Comparison of MAP and Posterior Mean Estimators

LEAP estimates liabilities by finding the maximum a posteriori (MAP) estimate of genetic effects for causal variants. Another possible estimator is the posterior mean estimator (PME), defined as

$$\hat{l} = E_{\beta,e}[X\beta + e \mid X, p; K]$$

where $X$ is a matrix of genotyped variants, $p$ is a binary vector of observed case-control phenotypes, $\beta$ is a vector of genetic effects, $e$ is a vector of iid environmental terms and $K$ is the trait prevalence. While the PME cannot be evaluated analytically, it can be evaluated numerically via Gibbs sampling[11, 29]. Here we discuss theoretical considerations that affect the choice of estimator, and provide an empirical comparison of the two estimators under a variety of scenarios. Our empirical evaluation demonstrates that the MAP estimator employed by LEAP is more robust to the presence of population structure than the PME. For completeness, a Gibbs sampler specification is given below.

From a theoretical perspective, both the PME and the MAP estimator employed by LEAP (also termed posterior mode estimator) are point estimates of the liabilities vector. According to Bayesian decision theory, the choice of a point estimate is dictated by the underlying loss function[30]. PMEs minimize the mean square error (MSE) loss function, whereas MAP estimators minimize a binary loss function that is equal to one if the estimated parameters are equal to the true parameters, and zero otherwise. Both estimators are consistent and asymptotically unbiased.

While Bayesian decision theory can guide the choice of estimators in the presence of a known probabilistic model, real data often deviates from the underlying model assumptions. Recall that according to the liability threshold model, the liability for individual $i$ is given by $l_i = g_i + e_i$, where $g_i = \sum_{j=1}^{m} v_{ij}^* \beta_j$ is the genetic component of the liability, $\beta_j$ is the effect size of variant $j$, and $v_{ij}^*$ is the value of variant $j$ for individual $i$, standardized to have zero mean and unit variance. The true model holds only when all variants are correctly standardized, where the standardized value $v_{ij}^*$ of an observed variant $v_{ij}$ is given by $v_{ij}^* = \frac{v_{ij} - E[v_j]}{\sqrt{Var[v_j]}}$, and $E[v_j]$ and $Var[v_j]$ are the population mean and variance of variant $j$, respetively. Otherwise, the liability does not follow a standard normal distribution.

When a sample consists of a single homogeneous population, the mean and variance of every variant are well defined, and can often be taken from the literature. However, GWAS samples are often confounded due to population structure. In such scenarios, the correct values depend on the composition of populations in the sample. It is difficult to empirically estimate the mean and variance of all variants from a given sample, because (a) these values are biased under ascertainment, and (b) even if a sample is randomly ascertained, the number of variants is typically substantially larger than the sample size, leading to a large variance in the estimates of their mean and variance.

Another challenging aspect of empirical estimation of the mean and variance of genetic variants, is that the resulting sample covariance matrix (SCM) $\frac{1}{m}XX^T$ is singular (where X is a design matrix with standardized columns), because one degree of freedom is lost by zero-centering all columns of $X$. The PME described in ref.[29] uses the inverse of this matrix, and thus cannot be used in combination with standardization based on empirical estimates of the mean and variance. To overcome this problem, we regularized the SCM via the well-known Ledoit-Wolf (LW) formula[31]. Briefly, given a SCM $C$, the corresponding regularized covariance matrix $C^R$ is given by

$$C^R = (1-\alpha)C + \alpha \frac{Tr[C]}{m} I$$

where $Tr[C]$ is the trace of $C$, $m$ is the number of variants, $I$ is the identity matrix and $\alpha$ is the optimal shrinkage coefficient that minimizes the mean squared error between the estimated and true covariance matrix, whose determination is described by Ledoit and Wolf[31]. We note that LEAP also uses a covariance matrix to efficiently compute the eigenvalues of the matrix $X$ (Online Methods), but unlike the PME, it can readily use the empirical SCM. All experiments described in the main text use the empirical SCM.

We conducted empirical simulations to evaluate the performance of the MAP estimator employed by LEAP and the PME described in ref.[29] under a wide range of settings. The data simulation procedure was the same as in all other experiments. We evaluated the methods in the presence of a trait with 0.1% prevalence, three levels of population structure (corresponding to $F_{ST}$ values of 0, 0.01 and 0.05) and three levels of family relatedness (corresponding to fraction of sib-pairs of 0%, 3% and 30%). Ten data sets were generated for each combination of settings, with 6,000 individuals in each data set. The true underlying heritability level of 50% was assumed to be known in all simulations, to prevent evaluation bias due to misestimating the heritability. The PME was invoked with 10,000 burn-in and 20,000 Gibbs sampler iterations (using 10,000 iterations yielded effectively the same results). Evaluation of the accuracy of liability estimates was only applied for controls, because liabilities of cases are trivial to find for rare diseases, as they are tightly clustered near the liability cutoff (Supplementary Fig. S1).

LEAP was evaluated using both the empirical, the LW regularized and the true covariance matrices. The PME was evaluated using only the LW regularized and the true covariance matrices, because it cannot be used with the empirical SCM due to the singularity noted above. The true covariance matrix was computed by standardizing each SNP according to its true mean and variance. These values were computed according to the appropriate formulas for a binomially distributed random variable with two trials (corresponding to the number of minor alleles) and success probability corresponding to the minor allele frequency (MAF). In experiments with population structure, we used the MAF corresponding to the mean MAF of the two populations.

To prevent biased liability estimation due to family relatedness, we employed a greedy algorithm to find the smallest subset of individuals whose removal leaves no pair of individuals with correlation coefficient greater than 0.05. The liabilities for individuals in this subset were estimated only after fitting the models without considering these individuals, as described in the online methods and in ref.[29].

The results demonstrate that when there is no population structure, liability estimation accuracy is greatest when the true covariance matrix is known (Supplementary Table S4). In such cases, LEAP and the PME were equally accurate. When the true covariance matrix was unknown, LEAP was always as or more accurate than the PME. In the presence of population structure, use of LEAP with the empirical SCM always lead to the greatest accuracy. We note that ref.[29] proposed dealing with population structure by regressing out top principal components from the genotypes while adjusting the liability threshold appropriately, but this procedure reduced accuracy for both methods (results not shown).

We conclude that covariance estimation in confounded and ascertained case-control studies is a challenging open problem that can substantially affect liability estimation. Our experiments demonstrate that the MAP estimator employed by LEAP is more robust to model violations than the PME described in ref.[29], and is thus more suitable for liability estimation in the presence of confounding. Another advantage of this estimator is its substantially reduced computational cost. We emphasize that although the PME used in ref.[29] cannot use the empirical SCM, potential alternative formulations may not suffer from this limitation. Nevertheless, because MAP results are similar whether the empirical or the LW SCM is used, this similarity is likely to hold under PMEs as well.

**Supplementary Table S4**: Comparison of accuracy of liability estimators, for controls only. The evaluated methods are the MAP estimator employed by LEAP, and the posterior mean estimator described in ref.[29]. The measures shown are the Pearson correlation between the estimated and the true liabilities, and the root mean square error. For each measure, the table shows the average, minimum, and maximum values obtained across 10 simulated data sets. Both evaluated methods were invoked with both the Ledoit-Wolf (LW) regularized and the true covariance matrix. The MAP estimator was additionally invoked with the empirical sample covariance matrix. Results for methods that obtained an average Pearson correlation at least 0.02 greater than the alternative method in the same setting, when using the same covariance matrix, are highlighted in bold.

| $F_{ST}$ | %sibs | Covariance | Measure | MAP | Posterior Mean |
|---|---|---|---|---|---|
| 0 | 0% | Empirical | Correlation | 0.26 (0.24-0.29) | - |
| | | | RMSE | 1.22 (1.19-1.24) | - |
| | | LW | Correlation | **0.26 (0.23-0.29)** | 0.10 (0.07-0.12) |
| | | | RMSE | **1.22 (1.19-1.24)** | 1.34 (1.32-1.37) |
| | | True | Correlation | 0.33 (0.31-0.36) | 0.33 (0.31-0.36) |
| | | | RMSE | 1.16 (1.14-1.18) | 1.16 (1.14-1.18) |
| | 3% | Empirical | Correlation | 0.25 (0.23-0.28) | - |
| | | | RMSE | 1.22 (1.2-1.24) | - |
| | | LW | Correlation | **0.25 (0.22-0.27)** | 0.21 (0.17-0.24) |
| | | | RMSE | **1.23 (1.21-1.25)** | 1.26 (1.23-1.29) |
| | | True | Correlation | 0.32 (0.29-0.34) | 0.32 (0.28-0.34) |
| | | | RMSE | 1.17 (1.15-1.20) | 1.17 (1.15-1.20) |
| 0 | 30% | Empirical | Correlation | 0.33 (0.3-0.36) | - |
| | | | RMSE | 1.16 (1.13-1.18) | - |
| | | LW | Correlation | **0.33 (0.3-0.36)** | 0.10 (0.07-0.15) |
| | | | RMSE | **1.16 (1.13-1.18)** | 1.34 (1.30-1.36) |
| | | True | Correlation | 0.38 (0.34-0.41) | 0.38 (0.34-0.41) |
| | | | RMSE | 1.11 (1.08-1.15) | 1.11 (1.08-1.15) |
| 0.01 | 0% | Empirical | Correlation | 0.21 (0.16-0.26) | - |
| | | | RMSE | 1.26 (1.22-1.30) | - |
| | | LW | Correlation | 0.19 (0.14-0.25) | 0.19 (0.14-0.25) |
| | | | RMSE | 1.27 (1.22-1.32) | 1.27 (1.22-1.32) |
| | | True | Correlation | 0.19 (0.15-0.22) | 0.19 (0.15-0.23) |
| | | | RMSE | 1.27 (1.25-1.31) | 1.27 (1.24-1.30) |
| | 3% | Empirical | Correlation | 0.23 (0.2-0.26) | - |
| | | | RMSE | 1.24 (1.22-1.26) | - |
| | | LW | Correlation | **0.21 (0.18-0.25)** | 0.19 (0.16-0.22) |
| | | | RMSE | **1.26 (1.23-1.28)** | 1.28 (1.25-1.30) |
| | | True | Correlation | 0.20 (0.16-0.24) | 0.21 (0.17-0.26) |
| | | | RMSE | 1.26 (1.23-1.3) | 1.25 (1.22-1.29) |
| | 30% | Empirical | Correlation | 0.37 (0.32-0.42) | - |
| | | | RMSE | 1.12 (1.08-1.17) | - |
| | | LW | Correlation | **0.36 (0.32-0.4)** | 0.24 (0.13-0.37) |
| | | | RMSE | **1.13 (1.09-1.17)** | 1.23 (1.12-132) |
| | | True | Correlation | 0.33 (0.25-0.42) | **0.35 (0.30-0.43)** |
| | | | RMSE | 1.16 (1.08-1.22) | **1.14 (1.07-1.18)** |
| 0.05 | 0% | Empirical | Correlation | 0.24 (0.14-0.45) | - |
| | | | RMSE | 1.24 (1.05-1.31) | - |
| | | LW | Correlation | 0.23 (0.12-0.45) | 0.22 (0.12-0.45) |
| | | | RMSE | 1.24 (1.05-1.32) | 1.24 (1.05-1.32) |
| | | True | Correlation | 0.23 (0.14-0.46) | 0.23 (0.14-0.46) |
| | | | RMSE | 1.24 (1.04-1.31) | 1.24 (1.04-1.31) |
| | 3% | Empirical | Correlation | 0.21 (0.15-0.26) | - |
| | | | RMSE | 1.26 (1.22-1.30) | - |
| | | LW | Correlation | 0.20 (0.15-0.25) | 0.20 (0.15-0.25) |
| | | | RMSE | 1.26 (1.23-1.31) | 1.27 (1.23-1.31) |
| | | True | Correlation | 0.20 (0.13-0.25) | 0.20 (0.13-0.25) |
| | | | RMSE | 1.26 (1.23-1.32) | 1.26 (1.23-1.32) |
| | 30% | Empirical | Correlation | 0.25 (0.13-0.40) | - |
| | | | RMSE | 1.23 (1.09-1.32) | - |
| | | LW | Correlation | 0.25 (0.14-0.40) | 0.25 (0.14-0.40) |
| | | | RMSE | 1.22 (1.09-1.31) | 1.22 (1.09-1.31) |
| | | True | Correlation | 0.24 (0.15-0.40) | 0.25 (0.15-0.40) |
| | | | RMSE | 1.23 (1.1-1.31) | 1.22 (1.10-1.30) |

In the remainder of this section, we describe the Gibbs sampler used to estimate the posterior mean of the liabilities. This posterior mean can be estimated by drawing $C$ pairs of samples $\beta^{(c)}, e^{(c)}$ from their posterior distribution for $c = 1, 2, \ldots, C$, and then estimating $\hat{l}$ via

$$\hat{l} \approx \frac{1}{C} \sum_{c=1}^{C} X\beta^{(c)} + e^{(c)}$$

Golan and Rosset[29] and Hayeck et al.[11] both recently described a sampling scheme for $\beta^{(c)}, e^{(c)}$ via a Gibbs sampler. We note that while the formulation described here (in notation different from these previous works) is mathematically equivalent to the sampler of Golan and Rosset[29], there are substantial differences in the implementation details which can lead to different behaviour in practice, as discussed above.

Gibbs samplers require specifying the posterior distribution of every sampled random variable, conditional on all the other random variable. We therefore derive the posterior distribution of each genetic effect $\beta_j$ for variant $j$, and each environmental term $e_i$ for individual $i$, and demonstrate that each of these follows a truncated normal distribution, which can be sampled from efficiently.

Denoting $\beta_{-j}$ as the effect sizes of all variants other than $j$, $t = \Phi^{-1}(1 - K)$ as the liability cutoff, encoding $p_i = 1$ for controls and $p_i = -1$ for cases, and assuming $\beta_j \sim N(0; \sigma^2)$, the posterior distribution of $\beta_j$ is given by

$$
\begin{aligned}
P(\beta_j \mid \beta_{-j}, e, X, p\,; K, \sigma^2) &= P(\beta_j;\, \sigma^2) \frac{P(p \mid \beta, e, X\,; K)}{P(p \mid \beta_{-j}, e, X\,; K)} \\
&= P(\beta_j;\, \sigma^2) \frac{I[p(X\beta + e) \leq pt]}{\int P(\tilde{\beta}_j;\, \sigma^2) P(p \mid \tilde{\beta}, e, X\,; K) d\tilde{\beta}_j} \\
&= P(\beta_j;\, \sigma^2) \frac{I[p(X\beta + e) \leq pt]}{\int P(\tilde{\beta}_j;\, \sigma^2) I[p(X\tilde{\beta} + e) \leq pt] d\tilde{\beta}_j} \\
&= P(\beta_j;\, \sigma^2) \frac{I[\beta_j \in B]}{\int_B P(\tilde{\beta}_j;\, \sigma^2) d\tilde{\beta}_j}
\end{aligned}
$$

Here, the inequality $p(X\beta + e) \leq pt$ is evaluated component-wise, $\tilde{\beta}$ is the vector of effect sizes composed of $\beta_{-j}$ and $\tilde{\beta}_j$, and $B$ is the subspace of $\tilde{\beta}_j$ wherein $p(X\tilde{\beta} + e) \leq pt$ holds. This is the definition of the probability density of a truncated normal distribution, and so we conclude that $\beta_j$ conditional on all other variables follows a truncated normal distribution.

Using similar notations, the posterior distribution for $e_i$ is given by

$$
\begin{aligned}
P(e_i \mid \beta, e_{-i}, X, p\,; K, \sigma_e^2) &= P(e_i;\, \sigma_e^2) \frac{P(p \mid \beta, e, X\,; K)}{P(p \mid \beta, e_{-i}, X\,; K)} \\
&= P(e_i;\, \sigma_e^2) \frac{I[p(X\beta + e) \leq pt]}{\int P(\tilde{e}_i;\, \sigma_e^2) P(p \mid \beta, \tilde{e}, X\,; K) d\tilde{e}_i} \\
&= P(e_i;\, \sigma_e^2) \frac{I[e_i \in A]}{\int_A P(\tilde{e}_i;\, \sigma_e^2) d\tilde{e}_i}
\end{aligned}
$$

where $A$ is the subspace of $\tilde{e}_i$ wherein $p(X\beta + \tilde{e}) \leq pt$ holds. As before, we conclude that $e_i$ conditional on all other variables follows a truncated normal distribution.

## Runtime Performance

We implemented the Probit model employed in LEAP in a custom Python package, using the scipy.optimize package[32]. On a Linux workstation with an Intel Xeon 2.90GHz CPU using a single core, the computation for a data set with 8,000 individuals takes less than five minutes. We note that the computation time is independent of the number of SNPs, because it uses the eigendecomposition of the genetic similarity matrix, which is already computed by an LMM, and is thus available at no additional computational cost. The computations can be sped up by computing the Hessian matrix numerically rather than analytically. In this case, the computations typically take less than a minute, at a negligible loss of precision.

In contrast, the computation of the joint MAP (Supplementary Note 3) for such a data set typically ranges between 30 and 50 minutes (using the Mosek quadratic solver; http://www.mosek.com). The computation of the liabilities posterior mean using GeRSI[29], with 10,000 burn-in iterations and 20,000 sampling iterations, takes approximately 45 minutes.

The reported running times are the times needed to estimate liabilities for a single left-out chromosome. The effective running time for all methods should be multiplied by the number of chromosomes, because a different liability estimator has to be estimated for each left-out chromosome.

## Insensitivity to Ascertainment in LEAP

Case-control studies are typically ascertained, having a greater proportion of cases in the study than in the general population. The models presented in the methods section did not explicitly account for the case-control sampling scheme. However, we show here that such a correction is not needed for the models considered.

Using the notations previously presented, we consider a case-control phenotype vector $p$, a matrix of genetic variants $X$, a vector of effect sizes $\beta$, a vector of environmental effects $e$ and a disease with prevalence $K$.

We introduce the selection indicators vector $s$, where $s_i = 1$ indicates that individual $i$ was selected to participate in the study, and $s_i = 0$ otherwise. In practice, $s_i = 1$ for every observed individual, and thus the likelihood of the observed phenotypes is conditional on $s_i = 1$ for every individual. For simplicity, we use the notation $s = 1$ as a shorthand for $s_1 = 1, s_2 = 1, \ldots, s_n = 1$.

Our key assumption is that $s$ is conditionally independent of $\beta$ and $e$, given $X$ and $p$. This assumption captures many common ascertainment schemes, as described below. Under this assumption, we can write the joint MAP maximization problem as

$$max_{\beta,e} P(\beta, e \mid s = 1, p, X; K) = max_{\beta,e} P(\beta, e \mid p, X; K)$$

where we make use of the conditional independence between $S$ and $\beta, e$. Thus, the objective function is equivalent to that of a non-ascertained study. The derivation for the genetic MAP estimator is equivalent, with the exception that $e$ is integrated out, as described in the main text.

The posterior mean liabilities are also unaffected by case-control sampling when the conditional independence holds. The posterior mean can be written as $E[X\beta + e \mid p, s = 1, X; K]$. Due to the conditional independence, this quantity is equal to $E[X\beta + e \mid p, X; K]$, and thus ascertainment does not need to be considered here either.

We caution that insensitivity to ascertainment does not imply that the fitted model is equivalent to a model that would have been fitted under a non-ascertained study. Rather, insensitivity to ascertainment means that given an ascertained study, a model that is fitted without taking the ascertainment procedure explicitly into account is equivalent to one that explicitly models the ascertainment scheme.

We now consider several different ascertainment schemes, and demonstrate that they obey the conditional independence assumption. Our derivation make use of graphical models theory[33], which is a rich language for expressing and inferring conditional independencies. Specifically, we consider probabilistic directed acyclic graphical models, wherein nodes represent random variables and parameters. According to the theory of probabilistic graphical models, two sets of random variables $Q_1$, $Q_2$ are conditionally independent given a set of random variables $W$, if no path in the graph between a node in $Q_1$ and a node in $Q_2$ obeys the d-connection criterion. A path obeys this criterion if (a) every node on the path with two incoming arcs is in $W$ or has a descendant in $W$, and (b) there are no nodes on the path that are in $W$ but do not have two incoming arcs on this path.

We consider three common ascertainment schemes, derive the appropriate graphical model for each one, and demonstrate that the conditional independence holds.

The first ascertainment scheme selects individuals to participate in the study based on their phenotype alone. This ascertainment scheme corresponds to Fig. S16a, wherein $s$ has a single incoming arc from $p$. It is easy to verify that paths between $\beta$ and nodes in $s$ do not obey the d-connection criterion.

The second ascertainment scheme selects individuals to participate in the study based on their phenotype and genetic ancestry. This can be encoded by introducing a population variable $r$ that governs the genotypes distribution, as shown in Fig. S16b. As before, no path between $\beta$ and $s$ follows the d-connection criterion, indicating conditional independence.

The third ascertainment scheme selects individuals to participate in the study based on their phenotypic and genotypic values. While less common, this ascertainment scheme may become more common in the future, as large panels will allow matching of cases with similar controls. The resulting model, shown in Fig. S16c, indicates that the conditional independence assumption holds.

Finally, we note that it is easy to verify that the conditional independence assumption holds under a wide variety of different assumptions (e.g., a population-specific prevalence, or when the prevalence is treated as an unknown random variable).

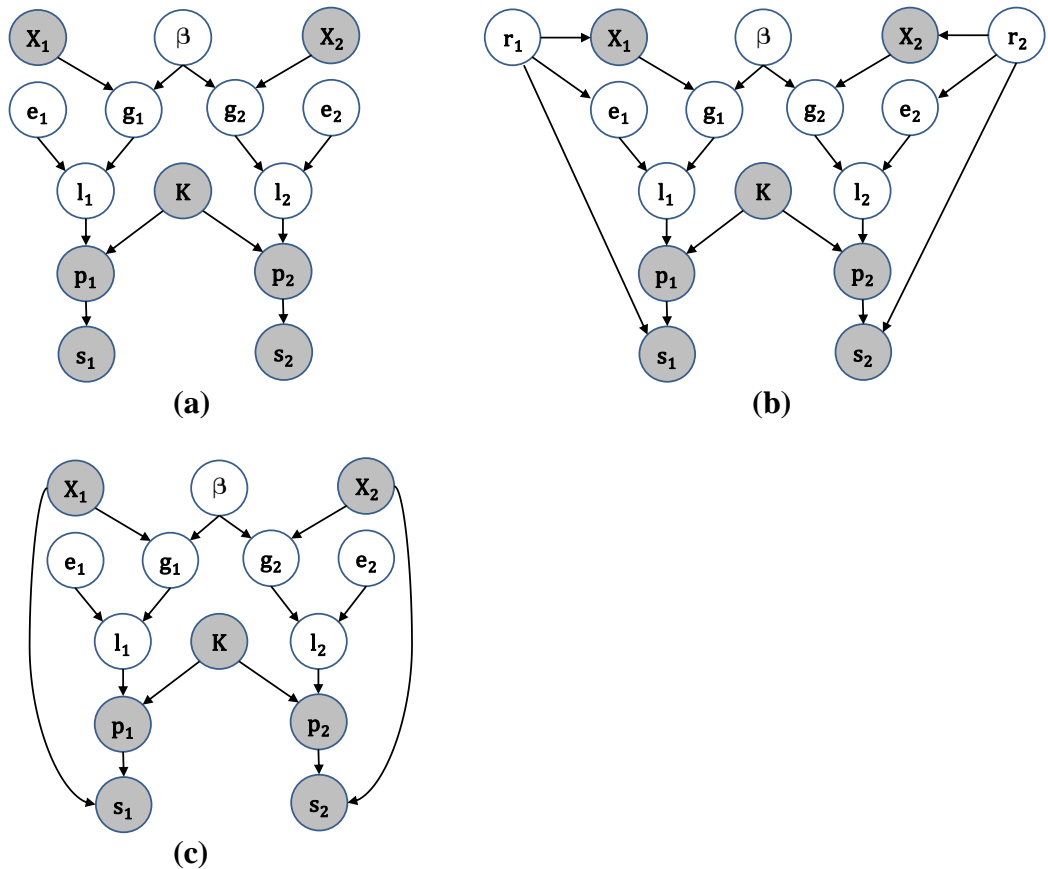

**Supplementary Figure S16:** Probabilistic graphical models for the disease model considered in the text. Each node corresponds to a parameter or to a random variable, using the same notations as in the text. Gray shaded nodes indicate parameters or conditioned variables whose values are known. Each panel shows a different ascertainment scheme. In all panels, the model shown is for a data set with 2 individuals. Panel (a) considers an ascertainment scheme that is determined only according to the case-control phenotype. Panel (b) considers an ascertainment scheme that considers both the phenotype and the genetic ancestry of an individual. Panel (c) considers an ascertainment scheme that considers both the phenotype and genotype of an individual. In all cases, $\beta$ is conditionally independent of the selection variables $s_1$ and $s_2$, given the genotypes and the phenotypes.

## Investigating the Causes of Power Loss in Ascertained Case Control Studies

Here we investigate several possible causes for power loss in LMMs under ascertained case-control studies. As discussed in the main text, it has recently been discovered that LMM performance in such studies deteriorates with increasing sample size, leading to loss of power compared to alternative methods[7]. Yang *et al.* have demonstrated that the severity of power loss in LMMs increases with the ratio between the sample size and the effective number of common genetic variants used to estimate kinship[7]. This ratio increases with sample size, regardless of genotyping density, because the effective number of common variants is bounded and is smaller than the number of genotyped variants, owing to linkage disequilibrium[26]. Thus, although the absolute power of LMMs increases with sample size, the increase is expected to be small compared to alternative methods, owing to model misspecification.

The loss of power of LMMs in ascertained case-control studies stems from violation of several of their modelling assumptions. First, LMMs assume that variants have an additive effect on the phenotype, which is obviously not true for case-control phenotypes. Second, LMMs assume that genetic and environmental disease factors are independent. However, it has recently been demonstrated that these factors become correlated under ascertainment, leading to a biased estimation of the genetic variance of phenotypes[2]. Third, LMMs assume that variants used to estimate kinship are independent of tested variants. However, several recent studies have demonstrated that causal variants tend to become correlated under ascertainment, because cases of rare diseases are likely to carry high dosages of risk alleles in multiple causal variants[7, 9, 10, 34-36]. The first violation is unique to case-control phenotypes, whereas the other two are also encountered in ascertained studies of continuous phenotypes.

To assess the influence of the first model violation on power, we considered an LMM which tests for association with the true underlying liabilities. To assess the influence of the second violation on power, we considered a "true $h^2$ LMM" in which the LMM parameter $\delta$ was fitted according to the true underlying narrow-sense heritability $h^2$ via the formula $\delta = 1/h^2 - 1$. This solution applies only to studies of continuous phenotypes, because additive narrow-sense heritability is not well defined for case-control phenotypes. We thus applied this solution only in conjunction with the first solution of liability-aware LMMs. We point out that an LMM with knowledge of both the true heritability and of the true liabilities represents an idealized version of LEAP, where both quantities were perfectly estimated.

To assess the influence of the third violation on power, we considered an "Oracle LMM" which estimates kinship out of all variants except for the causal ones. It is impossible to utilize such an LMM in practice, because the true identities of all true causal SNPs are unknown. We also considered oracle LMMs with knowledge of the true liabilities. We note that oracle LMMs cannot be used with knowledge of the true heritability. This is because the true heritability is the proportion to which causal variants in the kinship matrix influence the phenotypic variance, whereas there are no causal variants in the oracle LMM kinship matrix. We therefore fit the value of $\delta$ for oracle LMMs by maximizing the restricted maximum likelihood of the null model, as is common in LMM usage[24] (the estimated heritability in such cases is different from 0, because causal and non-causal variants are confounded due to the presence of population structure and family relatedness).

We evaluated data sets with sample sizes of 2,000, 4,000, 6,000 and 8,000 individuals, generated as described in the main text, with regular default parameters. Specifically, disease prevalence was 0.1% and heritability was 50%. For completeness, we also evaluated the performance of linear regression models, both with and without using the top 10 principal components as covariates.

The results are shown in Fig. S17. The largest improvements came from simply using the true liabilities, indicating that violations two and three have only a minor effect on power loss in continuous phenotype studies. Additional but smaller improvements came from use of the oracle LMM and use of the true heritability, along with the true liabilities. We conclude that while it is difficult to tease apart a single source of power loss, it is clear that liability estimation can greatly mitigate such loss.

We note that LEAP can be regarded as a noisy version of an LMM with knowledge of both the true liabilities and the true heritability, which was the best performing method. With increased sample sizes, it is expected that both heritability and liabilities can be estimated more accurately, enabling LEAP to more closely resemble such an LMM. Thus, the advantage of LEAP over a standard LMM is expected to increase with sample size, as demonstrated by the experiments in the main text.

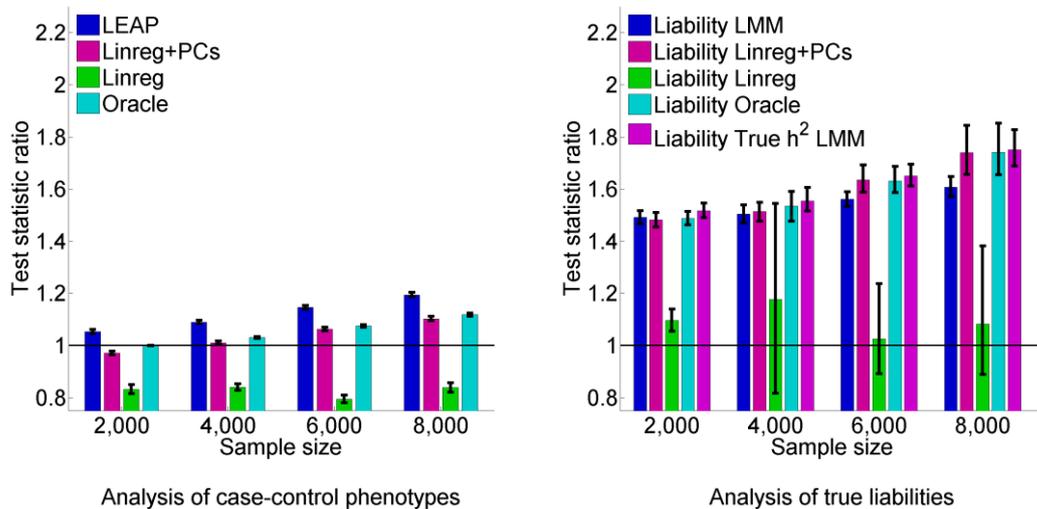

**Supplementary Figure S17:** Evaluations of factors that drive power loss in ascertained studies. All methods were compared to a standard LMM with no knowledge of the true causal SNPs, the underlying heritability or the true liabilities.

### Complex Ascertainment Schemes

The experiments described in the main text use a balanced ascertainment scheme and uniform trait prevalence, which encode the assumptions that individuals from different populations have an equal probability of being sampled, and that the studied trait has the same prevalence under different populations, respectively. Here we examine the sensitivity of LEAP to deviation from these assumptions.

We first examined scenarios wherein the ascertainment scheme differs between two populations. Specifically, we considered samples where a pre-specified number of cases is drawn from one population, and a different number of cases is drawn from the second population, such that the total number of individuals from each population is equal to 3,000. We evaluated the performance of LEAP under various levels of ascertainment imbalance. Our results indicate that the performance of LEAP improves as ascertainment imbalance increases, because the genotypes distribution among cases and controls becomes increasingly differentiated, leading to more accurate liability estimates (Fig. S18).

In another experiment, we evaluated the performance of LEAP in the presence of two populations with a different prevalence of the studied trait. We simulated data sets with two populations having trait prevalence of 0.1% and 1%, respectively, with an equal number of cases and controls in each population, and an equal number of individuals sampled from each population. LEAP was invoked by specifying the trait prevalence as either 0.1%, 0.5% or 1% in the heritability estimation and model fitting stages. The performance of LEAP was highly similar in all cases, with an average increase in the normalized test statistic of causal SNPs over a standard LMM of 7.8%, 7.9% and 7.9%, respectively. These results indicate that LEAP is robust to prevalence misspecification.

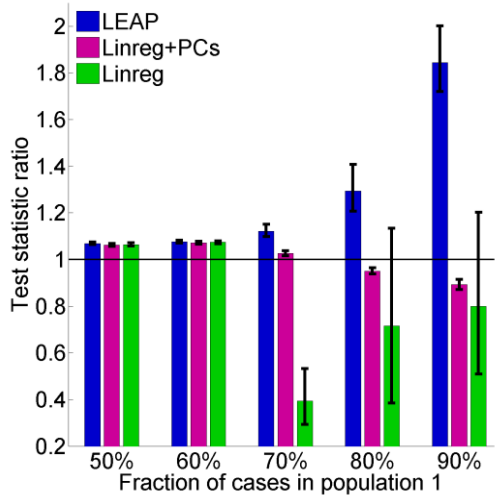

**Supplementary Figure S18**: Evaluation of GWAS performance in the presence of an imbalanced ascertainment scheme, wherein the majority of cases belong to population 1.

## Comparing LEAP with previous works

LEAP bears similarities to several recently developed methods for estimating the portion of the liability that is explained by a given set of explanatory variables[9, 10]. These methods are designed to prevent power loss in ascertained case-control studies using covariates, which arises due to induced correlations between tested and conditioned variables, as discussed above. However, these methods estimate the liability explained by a small set of covariates, whereas LMMs implicitly condition association tests on the entire genome, owing to the well known equivalence between LMMs and linear regression[37, 38]. A second key difference is that the aforementioned methods test variants for association with the residuals of the estimated liabilities, after regressing out the influence of the explanatory variables. In contrast, LEAP directly tests variants for association with the estimated liabilities, while effectively conditioning on all genome-wide variants. The use of genome-wide variants in both the liability estimation and in the association testing stage prevents spurious results due to confounding. We empirically verified that testing for association with estimated liabilities with either Linreg or Linreg+PCs, or with the estimated residuals via LEAP, leads to test statistic inflation (results not shown).

In parallel work, Hayeck et al. proposed an alternative framework, called LTMLM, for association testing under ascertained case-control sampling in the presence of confounding[11]. Both LTMLM and LEAP first estimate latent liability values and then test for association with these estimates. However, LTMLM tests for association with the posterior mean of the liabilities in a score test framework, whereas LEAP tests for association with the maximum a posteriori (MAP) of the liabilities. We found that utilizing the MAP estimator resulted in improved accuracy over the posterior mean estimator under a wide range of scenarios and at a substantially reduced computational cost (Methods and Supplementary Note 1).

**Accuracy of Liability Estimation**

We demonstrate here that estimation of effect sizes of genetic variants becomes increasingly accurate under increasing ascertainment, leading to increasing accuracy of liability estimates. We use the probabilistic model derived in the previous sections, and consider the MAP of the genetic effects vector $\beta$. We show that in balanced case-control studies, the likelihood function becomes increasingly sharply peaked around the MAP of $\beta$ with decreasing prevalence (and consequently, with increasing ascertainment).

According to standard statistical theory, the sharpness of the likelihood function is evaluated via the variance of its score at the MAP of $\beta$. The score is the gradient with respect to $\beta$ of the log-likelihood that is maximized in the Probit model. The variance of the score becomes equivalent to the Fisher information when there is no ascertainment and when $\beta$ is a fixed (unknown) parameter with no prior distribution. A higher score variance indicates that the likelihood function is more sharply peaked at the MAP of $\beta$, enabling more accurate estimation of $\beta$. The variance of the score is computed according to the true generative model. This model is similar to the Probit model, with the exception that $\beta$ is integrated out and the ascertainment procedure is taken into account.

Using the same probabilistic model described in the main text and in the previous section, the score is defined as $\frac{\partial \log(P(\beta \mid p, X; K))}{\partial \beta}$. The log likelihood function is explicitly given by

$$\log(P(\beta \mid p, X; K)) = \sum_{i \in controls} \log \Phi(t - X_i^T \beta; 0, \sigma_e^2) + \sum_{i \in cases} \log\left(1 - \Phi(t - X_i^T \beta; 0, \sigma_e^2)\right) - \frac{1}{2\sigma_g^2/m} \sum_j \beta_j^2 + U$$

where $X_i^T$ is the vector of genetic variants of individual $i$, $\beta$ is a vector of effect sizes, $\sigma_e^2$ is the variance of the environmental component of the liability, $\sigma_g^2$ is the variance of the genetic component of the liability, $m$ is the number of genetic variants, $t = \Phi^{-1}(1-K)$ is the liability cutoff for a disease with prevalence $K$, and $U$ is a term that does not depend on $\beta$. The score is therefore given by

$$\frac{\partial \log(P(\beta \mid p, X; K))}{\partial \beta}$$
$$= -\sum_{i \in \text{controls}} \frac{\phi(t - X_i^T\beta; 0, \sigma_e^2)}{\Phi(t - X_i^T\beta; 0, \sigma_e^2)} X_i^T + \sum_{i \in \text{cases}} \frac{\phi(t - X_i^T\beta; 0, \sigma_e^2)}{1 - \Phi(t - X_i^T\beta; 0, \sigma_e^2)} X_i^T$$
$$- \frac{1}{\sigma_g^2/m} \sum_j \beta_j$$

The variance of the score is computed according to the true generative model, which is similar to the Probit likelihood function, with the exception that $\beta$ is integrated out and there is conditioning on $s = 1$, using the same definition of $s$ as in the previous sections. Therefore, the distribution of $p$ used in the variance computation is

$$P(p \mid X, s = 1; K) = \frac{P(p, s = 1 \mid X; K)}{P(s = 1 \mid X; K)}$$
$$= \frac{1}{P(s = 1 \mid X; K)} \int \left[ \phi(\beta; 0, \sigma_g^2 I) \prod_{i \in \text{controls}} \Phi(t - X_i^T\beta; 0, \sigma_e^2) s_0 \prod_{i \in \text{cases}} \left(1 - \Phi(t - X_i^T\beta; 0, \sigma_e^2)\right) s_1 \right] d\beta$$

where $s_0$ and $s_1$ are the probabilities of including a control or a case in the study, respectively. The quantity $\frac{1}{P(s=1 \mid X;K)}$ is a normalization constant ensuring that the probabilities sum to one.

To demonstrate the effects of ascertainment on the score variance, we created data sets with 100 controls, 100 cases and a single binary variant. This formulation facilitates variance computations, because every value of the vector $p$ can be summarized via the number of controls and cases carrying a risk allele. Consequently, the variance can be computed exactly using a quadratic (rather than an exponential) number of likelihood computations. The normalization constant is implicitly computed by scaling all probabilities to ensure they sum to one.

We generated data sets with different distributions of the risk allele among cases and controls. For each data set we computed the score at the empirical MAP of $\beta$, using $\sigma_g^2 = 0.001$. We used values of $s_1 = 1$ and $s_0 = \frac{K}{1-K}$, which yield an equal mean number of sampled cases and controls. The integral was numerically computed using 1,000 equally spaced samples in the range [-2.5, 2.5]. Table S5 shows the score variance for various risk alleles distributions and prevalence values, indicating that the score increases with decreasing prevalence (and therefore, with increasing ascertainment). Therefore, more accurate liability estimation is obtained under increased ascertainment.

**Supplementary Table S5:** The variance of the score in the presence of a single binary variant. A higher score variance indicates that the effect size can be estimated more accurately. The values $p_0$ and $p_1$ are the fraction of controls and cases carrying the risk allele, respectively. For every tested pair of values, we report the score variance and the MAP of the variant effect size.

| | | Score Variance | | | | Effect Size MAP | | | |
|---|---|---|---|---|---|---|---|---|---|
| $p_0$ | $p_1$ | Prevalence | | | | Prevalence | | | |
| | | 0.1% | 1% | 10% | 50% | 0.1% | 1% | 10% | 50% |
| 10% | 10% | 1100.88 | 528.32 | 227.58 | 142.91 | 0 | 0 | 0 | 0 |
| 10% | 20% | 1081.98 | 512.39 | 222.56 | 143.30 | 0.042 | 0.034 | 0.024 | 0.020 |
| 10% | 30% | 1080.09 | 507.69 | 220.98 | 143.54 | 0.075 | 0.060 | 0.043 | 0.035 |
| 10% | 40% | 1085.20 | 507.60 | 220.81 | 143.69 | 0.103 | 0.083 | 0.060 | 0.049 |
| 10% | 50% | 1094.77 | 510.25 | 221.46 | 143.79 | 0.130 | 0.104 | 0.075 | 0.062 |
| 20% | 10% | 1186.42 | 557.30 | 234.83 | 143.30 | -0.044 | -0.035 | -0.025 | -0.020 |
| 20% | 20% | 1148.47 | 536.68 | 228.84 | 143.38 | 0 | 0 | 0 | 0 |
| 20% | 30% | 1132.85 | 527.58 | 226.18 | 143.49 | 0.035 | 0.028 | 0.020 | 0.016 |
| 20% | 40% | 1128.43 | 524.21 | 225.16 | 143.59 | 0.066 | 0.053 | 0.038 | 0.031 |
| 20% | 50% | 1131.28 | 524.48 | 225.19 | 143.67 | 0.095 | 0.076 | 0.055 | 0.045 |
| 30% | 10% | 1224.99 | 569.45 | 237.80 | 143.54 | -0.080 | -0.063 | -0.045 | -0.035 |
| 30% | 20% | 1184.69 | 549.54 | 232.15 | 143.49 | -0.036 | -0.029 | -0.020 | -0.016 |
| 30% | 30% | 1164.43 | 539.32 | 229.23 | 143.52 | 0 | 0 | 0 | 0 |
| 30% | 40% | 1155.72 | 534.62 | 227.89 | 143.57 | 0.032 | 0.026 | 0.018 | 0.015 |
| 30% | 50% | 1155.06 | 533.75 | 227.62 | 143.62 | 0.062 | 0.050 | 0.036 | 0.029 |

# Supplementary Note 2 - Detailed Methods

## Real Data Processing

In the WTCCC1 data sets, the control group consisted of individuals from the UK Blood Service Control Group (NBS) and from the 1958 British birth cohort. SNPs were excluded from the study if they had minor allele frequency <5%, missingness rates >1%, a significantly different missingness rate between cases and controls, or a significant deviation from Hardy-Weinberg equilibrium among the controls group. Individuals were excluded from the analysis if they were in the WTCCC exclusion lists or if they had missingness rates>1%.

In the multiple sclerosis (MS) and ulcerative colitis (UC) data sets, we used the same data processing described in ref.[7] to ensure consistency. Briefly, UK controls and cases from both UK and non-UK were used. SNPs were removed with >0.5% missing data, p<0.01 for allele frequency difference between two control groups, p<0.05 for deviation from Hardy-Weinberg equilibrium, p<0.05 for differential missingness between cases and controls, or minor allele frequency<1%.

In all analyses, SNPs within 5M base pairs of the human leukocyte antigen (HLA) region were excluded, because they have large effect sizes and highly unusual linkage disequilibrium patterns, which can bias or exaggerate the results[39].

## Methods Evaluation

Both power and sensitivity to confounding had to be measured in all experiments, because not all the evaluated methods properly control for type I error. We used two measures for this task. For simulated data, we computed the empirical type I error rate associated with each P value, and then computed power as a function of the type I error rate. The type I error rate associated with a P value is defined as the proportion of non-causal SNPs with P value smaller than this P value. The average power of each method was computed by averaging the power corresponding to 1,000 equally spaced significance thresholds between 0 and 1 (in log space). Confidence intervals were obtained via 10,000 bootstrap samples of test statistics, where test statistics of causal and non-causal SNPs were sampled separately to preserve their number.

For both simulated and real data, we also used another measure that directly compares two methods of interest and provides easily interpretable results. To this end, we first normalized all test statistics by dividing them with the genomic control (GC) inflation factor[28] $\lambda_{GC}$, defined as the ratio of observed to expected median test statistic in $\chi^2$ space. Afterwards, the ratio between the normalized test statistics obtained by the two methods for each known causal variant was computed. The methods were compared based on the mean of the ratios. Both proposed evaluation approaches assess empirical power given the true type I error rate. However, the first measure simply counts the number of test statistics exceeding the significance cutoff (which is dependent on sample size), whereas the second one is sensitive to systematic differences in the distribution of such test statistics.

The second measure employed (the mean of the ratios) is similar to the ratio of the means, which is the measure used in refs.[9, 10]. However, the ratio of the means is often dominated by excessively large test statistics, whereas the mean of the ratios assigns equal importance to all variants. Additionally, this measure has an intuitive interpretation as the mean increase in test statistics of causal variants. Confidence intervals for this measure were obtained via 10,000 bootstrap samples of test statistics of causal SNPs.

SNPs having p>0.01 under all methods were discarded from this analysis, because they tended to bias the results, while in practice the results for such SNPs are meaningless. When not removing these SNPs, LEAP gained an unfair advantage in the analysis of real data sets, with a mean ratio of 1.14 for MS, and 1.16 for UC, because many such SNPs gain a slightly higher test statistic under LEAP. Such SNPs greatly bias the test statistics ratio, despite being meaningless in practice. The ratio of the means is relatively unaffected by the exclusion of such SNPs, because it is dominated by variants with large test statistics.

For the analysis of real data, we measured normalized test statistics for known associated SNPs. In the MS and UC data sets, we used the list of tag SNPs published in ref.[7]. In the analysis of WTCCC1 phenotypes we used best tags from the NHGRI catalog[17] (having the highest $r^2$ measure with the causal SNP) as a bronze standard. Associated SNPs without a tag SNP having $r^2$>0.5 or whose best tag was within 5M base pairs of another best tag with a higher statistic were discarded.

Due to the small number of SNPs with p<0.01 in real data sets, we also report the ratio of the means for real data sets, using all tag SNPs. We computed the P value of having a ratio of means greater than the actual ratio via permutation testing, by swapping the two test statistics computed for a SNP by two different methods one million times.

In the computation of type I error for real data sets (Supplementary Tables S2-S3), SNPs within 2M base pairs of a tag SNP were excluded from the analysis. However, all SNPs were used for the computation of the genomic inflation factor, to obtain measures comparable with previous publications.

In experiments with covariates, the covariates were not used as fixed effects, because this leads to substantial power loss in ascertained case-control studies[10, 34-36]. LEAP used covariates by including them in the liability estimation stage and then regressing their effect out of the estimated liabilities (Supplementary Note 2).

## Simulations Methodology

To validate our results, we generated synthetic data sets with varying numbers of individuals and with 60,100 SNPs that do not affect the phenotype, as well as 50-5000 SNPs affecting the phenotype. The number of SNPs was selected according to the estimated effective number of SNPs in contemporary GWAS samples[7]. The SNPs were not in linkage disequilibrium, because it has been shown that the distribution of genetic similarity matrices is not affected by its presence[26].

We simulated population structure via the Balding-Nichols model[27], wherein allele frequencies in the range [0.05, 0.5] were randomly drawn for an ancestral population, and frequencies for two subpopulations were drawn from a Beta distribution with parameters $f(1 - F_{ST})/F_{ST}$ and $(1 - f)(1 - F_{ST})/F_{ST}$, where $f$ is the minor allele frequency in the ancestral population. Additionally, each data set contained 100 unusually differentiated SNPs, with an allele frequency difference of 0.6 between the two subpopulations (where the allele frequencies for the first subpopulation were randomly drawn from [0, 0.4]), to simulate ancient population divergence[40].

Family relatedness was simulated by creating sib-pairs (by generating two parents and a set of children, assuming all generated SNPs are unlinked) in one of the two subpopulations, as in ref.[3]. To combine family relatedness with ascertainment, we first created ascertained data sets with no related individuals. We then randomly selected pairs of individuals from the ascertained data set and designated them as parents, by generating pairs of children for every pair of parents. Afterwards, the two parents were removed from the data set.

Phenotypes were generated using the liability threshold model. Causal SNPs were generated for every individual, with effect sizes drawn from a standard normal distribution. To simulate differences in the environmental component of the liability between the two populations, we also created a hidden causal variable that acts as a hidden SNP generated according to the Balding-Nichols model, with an effect size drawn from a zero-mean normal distribution with a standard deviation of 5. The genetic component of the liability generated for each individual was standardized, so that the liability follows a standard normal distribution. The liability cutoff was selected by sampling 3,000/K individuals and finding the 1-K percentile of the phenotypes, where K is the disease prevalence, as in ref.[9].

Unless otherwise stated, all data sets were created with 0.1% prevalence, 6,000 individuals, 50% cases, 50% heritability, $F_{ST}$=0.01, and with 30% of individuals in one of the two subpopulations who are sib-pairs. Each individual carried 500 causal SNPs with different allele distributions between the two populations according to the $F_{ST}$ level, one of which was hidden. These SNPs accounted for the genetic component of the liability of each individual, according to the specified heritability level. The environmental component of the liability for each individual was drawn from a zero-mean normal distribution, with variance selected to ensure that the liability variance is one.

### LMM analyses

LMM analyses were performed with FastLMM 2.07[24] using default settings unless otherwise stated. In analysis of simulated case-control phenotypes the LMM parameter $\delta$, which controls the ratio of residual to genetic variance, was fitted only to the null model for performance considerations. It has been shown that for typical human diseases this approximation often makes a very small difference in practice[39]. We empirically verified that fitting $\delta$ for every SNP yielded almost identical results (results not shown). When using liability estimates as phenotypes, $\delta$ was determined according to the liability heritability estimator $\hat{h}^2$, which was estimated using the method of ref[2], with the formula $\hat{\delta} = \frac{1}{\hat{h}^2} - 1$.

When testing synthetic data sets, tested SNPs were excluded from the genetic similarity matrix to prevent proximal contamination[7, 24, 25]. In the analysis of real data sets, when testing SNPs on a certain chromosome, this chromosome was excluded from the LMM genetic similarity matrix to prevent proximal contamination[24, 25].

**Variant Selection**

Several recent papers proposed improving the power of LMMs by estimating kinship via a subset of variants that account for a large fraction of the phenotype variance[25, 41, 42]. This strategy can work well for continuous phenotypes in limited situations[43]. However, under ascertained case-control sampling, this strategy renders the problem of power loss under LMMs even more severe, because it increases the ratio between the number of individuals and the number of variants used by the LMM. Larger values of this ratio increase the severity of power loss, as shown in ref.[7] and in the results below (Supplementary Fig. S19).

We performed a series of experiments to evaluate the effect of variant selection on LMM performance, using two variant selection methods that differ in their selection criteria. Both methods select a subset of SNPs to be used in the LMM genetic similarity matrix (GSM) by first ordering all SNPs according to their univariate linear regression P value, and then considering increasingly larger subsets of SNPs having the lowest P values. The first method is LMM-select-$\lambda_{GC}$, , which selects the smallest subset of SNPs that yield a local minimum of the genomic control inflation factor[25]. The second variant selection method is LMM-select-pred, which uses the subset of variants having the largest out-of-sample predictive power, measured according to the predictive log likelihood obtained via five-fold cross validation[41, 42]. Following the recommendation of ref.[41], we included the top five principal components, associated with the five largest eigenvalues, as fixed effects under this selection method. For both methods, we evaluated subsets of sizes [1, 5, 10, 15, 20, 30, 40, 50, 75, 100, 250, 500, 750, 1000, 2000, 3000, 4000, 5000, 6000, 7000, 8000, 9000, 10000, 20000, 30000, 40000, 50000, all SNPs].

The results demonstrate that under case-control sampling, LMMs using a subset of SNPs suffer from a loss of power, whose severity increases as prevalence decreases (Supplementary Fig. S19). LMM-select-pred performs poorly under all settings, possibly because it directly attempts to select a subset of variants accounting for as large as possible fraction of the phenotype variance, which in turn attenuates the signal explained by tested variants, leading to loss of power[7]. In contrast, the SNPs subset selected by LMM-select-$\lambda_{GC}$ accounts for a smaller fraction of the phenotype variance, which can even improve power under random ascertainment. We therefore opted to not follow this strategy in our experiments.

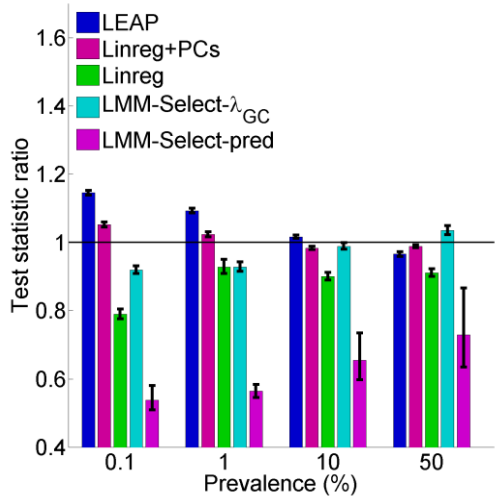

**Supplementary Figure S19**: Evaluation of variant selection methods. The figure shows the mean ratio of normalized test statistics for causal SNPs between each evaluated method and an LMM.

## Heritability Estimation

We estimated the heritability of the studied diseases using the method of ref.[2]. We excluded the top 10 principal components (PCs) from every correlation matrix prior to heritability estimation. In experiments with a covariate, the covariate was used as a fixed effect in the estimation stage.

We excluded individuals with at least one correlation coefficient >0.05 with another individual, because the method assumes that family relatedness is not present. We used a greedy algorithm where at each step the individual with the largest number of correlation coefficients >0.05 was removed. However, this procedure was not performed for experiments with extreme population structure ($F_{ST}$=0.05), because almost all individuals would be excluded, due to the fact that many SNPs have an extremely high frequency in one population compared to the other.

In the analysis of real data, a different heritability estimator was computed for every chromosome, because a different liability estimator was provided for every chromosome. This was done by considering only SNPs from other chromosomes in the computation. In the analysis of synthetic data, SNPs were randomly split into 10 pseudo-chromosomes, as described in the Liability Estimation section. Heritability was estimated for only one pseudo-chromosome, and this estimate was used for the liability estimation stage of all pseudo-chromosomes, because it is guaranteed that there is no systematic difference between the pseudo-chromosomes.

## Liability Estimation

Liabilities were estimated via a custom implementation of a regularized Probit model, described in detail below. For both synthetic and real data, liability estimation was performed on a per-chromosome basis. For every chromosome, we estimated the liability using only SNPs on other chromosomes. This is meant to guarantee that tested SNPs were not involved in the computation of the liability with which they are tested for association. For synthetic data sets, this was done by randomly dividing the SNPs into 10 pseudo-chromosomes. We note that typical GWAS also exclude SNPs on a given chromosome when testing for association to avoid proximal contamination[7, 25], which requires computing a different eigendeomposition for every chromosome. These eigendecomposions, which are also required for efficient computation of LEAP, are thus readily available at no further computational cost.

In the analysis of type I diabetes and rheumatoid arthritis, a single SNP from the HLA locus, having the lowest P value out of all SNPs, was used as a covariate in the liability estimation, owing to the strong influence of the HLA locus on these phenotypes.

**Probit Computations**

We estimated liabilities via a Probit model, by using the eigenvectors of the genotype matrix $X$ as the design matrix. The Probit model was fitted via Newton's method. This is an iterative fitting procedure that makes use of first and second order derivatives, defined as:

$$\beta^{i+1} \leftarrow \beta^i - \left[\frac{\partial^2 P(\beta, p \mid X; K)}{\partial \beta (\partial \beta)^T}\right]^{-1} \frac{\partial P(\beta, p \mid X; K)}{\partial \beta}$$

where $P(\beta, p \mid X; K)$ is the joint posterior density of the effect sizes $\beta$ and the phenotypes $p$. Denoting $Z$ as the matrix of eigenvalues of $X$, and $t$ as the liability cutoff, the first and second order derivatives are given by:

$$\frac{\partial P(\beta, p \mid X; K)}{\partial \beta} = \sum_i \left[p_i \frac{\phi(Z_i^T \beta - t)}{\Phi(Z_i^T \beta - t)} + (1 - p_i) \frac{\phi(Z_i^T \beta - t)}{1 - \Phi(Z_i^T \beta - t)}\right] Z_i - \frac{1}{\sigma_g^2/m} \beta.$$

$$\frac{\partial^2 P(\beta, p \mid X; K)}{\partial \beta (\partial \beta)^T}$$

$$= -\sum_i \phi(Z_i^T \beta - t) \left[p_i \frac{(Z_i^T \beta - t)/\sigma_e^2 \Phi(Z_i^T \beta - t) + \phi(Z_i^T \beta - t)}{\Phi^2(Z_i^T \beta - t)}\right.$$

$$\left. + (1 - p_i) \frac{-(Z_i^T \beta - t)/\sigma_e^2 \left(1 - \Phi(Z_i^T \beta - t)\right) + \phi(Z_i^T \beta - t)}{\left(1 - \Phi(Z_i^T \beta - t)\right)^2}\right] Z_i Z_i^T$$

$$- \frac{1}{\sigma_g^2/m}.$$

The initial values $\beta^0$ were selected by solving the L2-constrained linear regression problem $Z\beta = p$, using the same regularization parameter as in the Probit model.

When fitting the models, we excluded individuals having correlation coefficient >0.05 with at least one other individual, using the same greedy algorithm described earlier (except under extreme population structure, as described above). We used the fitted parameters to estimate liabilities for all individuals, including the excluded ones.

### Inclusion of Covariates

Inclusion of covariates in the LEAP framework presents both technical and statistical challenges. These challenges, and our proposed solutions, are described below.

A technical challenge arises because of numerical instabilities in naive use of Newton's method. Naively, Covariates can be included in the Probit model by adding additional columns to the design matrix Z, without adding corresponding terms to the penalty term of the Probit likelihood. However, standard application of Newton's method in this case can lead to numerical instabilities because of extreme differences in the scaling of columns of the Hessian matrix. To solve this, we perform an iterative gradient descent algorithm. At each iteration we fit the random and the fixed effects separately via Newton's method. When fitting the random effects (those affecting the penalty term) we treat the fixed effects as constants, and vice versa.

A statistical challenge is encountered because covariates and causal variants tend to become highly correlated in ascertained case-control studies. This induced correlation takes place because cases of rare diseases often carry high dosages of multiple risk variables[10, 34-36]. Furthermore, when fitting covariates as fixed effects, the analysis is no longer independent of the ascertainment scheme.

To exploit the information found in covariates without suffering from power loss and having to account for ascertainment, we include them as additional regularized random variables in the Probit model, using the same regularization strength as other variants. After obtaining a liability estimate, we regress the effect of the covariates out of the estimated liabilities, and test for association with this modified liability estimator. In the Probit model, the covariates are standardized to have zero mean and variance equal to the mean variance of all other variables used in the model.

We note that additional improvement may be gained by estimating the covariate effect size based on information from the literature[10].

### The Joint MAP Estimator

For estimation of the joint MAP of the genetic and environmental components of the liability (Supplementary Note 3), we used the MOSEK quadratic solver (http://www.mosek.com). Related individuals were handled in the same way as in LEAP. Namely, we first employed a greedy algorithm that excluded related individuals from the sample, as previously described. Afterwards, we applied the MAP estimator, which fitted a model of effect sizes $\beta$ and liability environmetal components $e$. Next, We estimated the genetic component $g$ of the excluded individuals via $g = X^r \beta$, where $X^r$ is the matrix of genotypes of excluded individuals. Finally, we determined the environmental component $e$ of the excluded individuals in the same way as LEAP. We also evaluated a model that did not exclude related individuals, which yielded slightly inferior results (results not shown).

## Supplementary Note 3 - LEAP Overview

To motivate the use of LEAP, we first introduce the liability threshold model and its relation to LMMs.

### The Liability Threshold Model

LEAP originates from the statistical framework of the liability threshold model[8], which is briefly presented here. A key assumption behind this model is that every individual $i$ carries a latent normally distributed liability variable $l_i \sim N(0,1)$. Cases are individuals whose liability exceeds a given cutoff $t$, i.e. $l_i \geq t$. The cutoff $t$ can be inferred given the disease prevalence $K$ as $t = \Phi^{-1}(1-K)$, where $\Phi^{-1}(\cdot)$ is the inverse cumulative probability density of the standard normal distribution.

The liability $l_i$ can be decomposed into two additive terms corresponding to the genetic and environmental effects affecting a trait, denoted as $g_i$ and $e_i$:

$$l_i = g_i + e_i. \quad (1)$$

Without loss of generalization, we assume that $g_i$ and $e_i$ are independently drawn from zero-mean normal distributions with variances $\sigma_g^2$ and $\sigma_e^2$, respectively, and thus $\sigma_g^2 + \sigma_e^2 = 1$. The genetic term $g_i$ for an individual is given by a linear combination of genetic variants and their corresponding effect sizes,

$$g_i = \sum_{j=1}^{m} v_{ij}\beta_j \quad (2)$$

where $\beta_j$ is the effect size of variant $j$ and $v_{ij}$ is the value of variant $j$ for individual $i$, standardized to have zero mean and unit variance. The effect sizes are assumed to be drawn iid from a normal distribution,

$$\beta_j \sim N(0, \sigma_g^2/m). \quad (3)$$

When the identities of truly causal variants are unknown, a commonly used assumption is that all genotyped variants have an effect size drawn from this normal distribution[2].

The genetic and environmental effects $g_i$ and $e_i$ are deeply related to the narrow-sense heritability[2] of a trait, defined as $h^2 = \sigma_g^2/(\sigma_g^2 + \sigma_e^2)$. This term is used to quantify the degree to which a given trait is affected by genetic factors. Recently, methods for estimating the underlying heritability of the liability of case-control traits have been proposed[2, 22]. These methods can be used to estimate the heritability, and consequently the variances $\sigma_g^2$ and $\sigma_e^2$.

## Linear Mixed Models

To motivate LEAP, we first present the LMM framework. For a given sample of individuals, LMMs assume that an observed phenotypes vector $y$ follows a multivariate normal distribution

$$y \sim N(\mu, \sigma_g^2 C + \sigma_e^2 I) \quad (4)$$

where $\mu$ is the distribution mean, $I$ is the identity matrix, $C$ is a covariance matrix encoding genetic correlations between individuals, and $\sigma_g^2, \sigma_e^2$ are the variances of the genetic and environmental components of the covariance, respectively. This model naturally encodes the assumption that genetically similar individuals are more likely to share similar phenotypes. The genetic covariance matrix $C$ is often estimated from genotypes variants as $C = \frac{1}{m} X X^T$, where $X$ is a design matrix of genotyped variants, standardized so that all columns have zero mean and unit variance. Association testing for a given variant $v$ can be carried out by assigning $\mu = \mu_0 + v\alpha_v$, where $\alpha_v$ is the variant effect size, and attempting to reject the null hypothesis $\alpha_v = 0$ by fitting the model via restricted maximum likelihood[44].

A close relation between the LMM and the liability threshold model is revealed by considering the relation between an LMM and linear regression, wherein effect sizes are drawn from $N(0, \sigma_g^2/m)$. Denoting $\varphi(y; \mu, \Sigma)$ as the probability density of the multivariate normal distribution, and using basic properties of the normal distribution, the LMM can be rewritten as follows.

$$\varphi(y; \mu, \sigma_g^2 C + \sigma_e^2 I) = \varphi\left(y; \mu, \frac{\sigma_g^2}{m} X X^T + \sigma_e^2 I\right) = \int \varphi(y; \mu + X\beta, \sigma_e^2 I) \, \varphi\left(\beta; 0, \frac{\sigma_g^2}{m} I\right) d\beta. \quad (5)$$

The phenotypes distribution under LMMs is therefore equivalent to the liability distribution under the liability threshold model, after integrating the effect sizes out.

This interpretation of LMMs provides a straightforward way to extend them to handle binary phenotypes. Given a disease with prevalence $K$ and a corresponding liability cutoff $t$, the likelihood for a given case-control status vector p conditional on K is given by

$$P(p \mid X; \mu, \sigma_g^2, \sigma_e^2, K) =$$

$$\int \left[ \varphi\left(\beta; 0, \frac{\sigma_g^2}{m} I\right) \prod_{i \in controls} \Phi(t - \mu - X_i^T \beta; 0, \sigma_e^2) \prod_{i \in cases} \left(1 - \Phi(t - \mu - X_i^T \beta; 0, \sigma_e^2)\right) \right] d\beta \quad (6)$$

where $\Phi(y; \mu, \Sigma)$ is the cumulative probability density of the normal distribution, and $X_i^T$ is the $i^{th}$ row of X. The relation to the liability threshold model can be made clearer by rewriting this likelihood as

$$P(p \mid X; \mu, \sigma_g^2, \sigma_e^2, K) = \int_V P(l \mid X) dl \qquad (7)$$

where $l = \mu + X\beta + e$ is the underlying liability, V is the subspace wherein $l_i \geq t$ for cases and $l_i < t$ for controls, and

$$P(l \mid X) = \int \varphi(l; \mu + X\beta, \sigma_e^2 I) \varphi\left(\beta; 0, \frac{\sigma_g^2}{m} I\right) d\beta = \varphi\left(l; \mu, \sigma_g^2 \frac{1}{m} XX^T + \sigma_e^2 I\right) \qquad (8)$$

is the liability density. Recall that $C = \frac{1}{m} XX^T$ is the LMM genetic covariance matrix. Thus, computing the likelihood of a case-control phenotype is equivalent to integrating the underlying liability over its support.

The above derivation suggests a natural way to perform association testing in the presence of case-control phenotypes. However, this requires fitting the parameters and performing a sampling procedure over liability values for every tested variant, resulting in excessively expensive computations that are infeasible in most circumstances.

## Accounting for Ascertainment in Association Testing

The previous section describes a model for computing the likelihood of a binary phenotype under the LMM framework. The presented model does not account for fixed effects or for ascertainment, where the proportion of cases in the study is greater than the disease prevalence. For completeness, we now derive a model that fully accounts for fixed effects and ascertainment as well. To this end, we use the selection variable $s$ introduced in the previous section, and consider a single covariate $v$ (e.g. a tested SNP) with a fixed effect $\alpha$. The extension to multiple covariates is straightforward. Using the same notations as before, the likelihood of an observed case-control phenotype, conditional on the genotypes, the fixed effect, the disease prevalence $K$ and on $s = 1$ is given by

$$P(p \mid v, X, s = 1; K, \alpha) = P(p \mid v, X; K, \alpha) \frac{P(s = 1 \mid p, X)}{P(s = 1 \mid v, X; K, \alpha)}$$

where we make use of the common assumption that $s$ is independent of all other variables given $p$ and X. The first quantity on the right hand side of this equation is the likelihood of the non-ascertained model, whose computation is intractable due to having to integrate over a high dimensional subspace of liabilities, as described in the main text.

An alternative formulation uses a retrospective approach, where the covariates are treated as random, and association testing is carried by computing the likelihood conditioned on the phenotype:

$$P(v, X \mid p, s = 1; K, \alpha) = P(v, X \mid p; K, \alpha)$$

where again we make use of the conditional independence assumption. Thus, under the retrospective likelihood approach, ascertainment does not need to be considered. By invoking Bayes rule, we obtain

$$P(v, X \mid p; K, \alpha) = P(p \mid v, X; K, \alpha) \frac{P(v, X; K, \alpha)}{P(p; K, \alpha)}$$

As before, the first term on the right hand side is the prospective likelihood, which is intractable to compute. Another drawback of this approach is that the genotypes distribution needs to be modeled, in contrast to the prospective approach.

## Liabilities Estimation

As discussed above, testing for associations under the liability threshold model requires integrating the underlying liability vector over its support. Motivated by this observation, we propose approximating such association testing by selecting a liability estimator and treating it as the observed phenotype vector. A good liability estimator has values close to the true, unobserved, underlying liability. Thus, the problem is equivalent to inferring the value of an unknown continuous variable with a known distribution.

Recall that the liabilities vector $l$ is given by $l = g + e$, where $g$ and $e$ are the genetic and environmental components of the liability, respectively. Further recall that $g$ is given by $g = X\beta$, where $X$ is the genotypes matrix and $\beta$ is a vector of effect sizes. We consider two closely related liability estimators: The joint maximum a posteriori estimate (MAP), and the genetic MAP. The first quantity jointly estimates the posterior mode of $\beta$ and $e$, conditional on the observed phenotypes, genotypes and the disease prevalence. The second quantity first estimates $\hat{\beta}$, the MAP of $\beta$, by considering $e$ as a nuisance parameter that is integrated out, and then finds the MAP of $e$ given $\hat{\beta}$. Although the first quantity has a clearer interpretation, the second quantity has favourable properties that render it superior in practice (detailed below), and will be used in LEAP.

Another natural estimator of $l$ is its posterior mean, which can be obtained via sampling[11, 29]. Our experiments have demonstrated that, in the presence of population structure, the genetic MAP estimator employed by LEAP obtains similar or greater liability estimation accuracy, at a significantly reduced computational cost (see Supplementary Note 1, Supplementary Fig. S20 and Supplementary Table S4 for a thorough investigation of this issue).

We now describe the derivation and computation of both MAP quantities in detail. Importantly, while the derivations below do not explicitly take the case-control sampling scheme into account, they yield identical results to derivations that do take the ascertainment procedure into account (Supplementary Note 1). Furthermore, while the optimization problems derived below are extremely high dimensional, they can readily be formulated as lower-dimensional problems with dimensionality equal to the sample size, as described in the next section. Inclusion of covariates is described in Supplementary Note 1.

### *Joint MAP Estimator*

The joint MAP maximizes the joint posterior likelihood of $\beta$ and $e$, conditional on the phenotypes, genotypes and the disease prevalence. Denoting $p$ as the vector of observed case-control phenotypes and $K$ as the disease prevalence, the likelihood to maximize can be written as

$$P(\beta, e \mid X, p; K) \propto P(\beta)P(e)P(p \mid \beta, e, X; K) \quad (9)$$

where the proportionality sign indicates that the likelihood is scaled by a constant that is independent of $\beta$ and $e$. Equation 9 makes use of the fact that $\beta$ and $e$ are marginally independent of $X, K$ and of each other. The probability $P(p \mid \beta, e, X; K)$ is a delta function that is equal to one if all cases/controls have liabilities greater/smaller than the cutoff $t = \Phi^{-1}(1-K)$, and zero otherwise. Therefore, using the definitions of $\beta$ and $e$, computing the MAP is equivalent to solving the optimization problem

$$argmax_{\beta,e}\ \varphi\left(\beta;\ 0, \frac{\sigma_g^2}{m}I\right)\varphi(e;\ 0,\sigma_e^2 I) \quad \text{s.t.}\quad p(X\beta + e) \leq pt \quad (10)$$

where we encode $p_i = 1$ for controls and $p_i = -1$ for cases, and the inequality is evaluated component-wise. Taking the logarithm, transforming the maximization to a minimization, and using the definition of the normal distribution, we obtain the equivalent problem:

$$argmin_{\beta,e}\ \frac{1}{2\sigma_g^2/m}\sum_j \beta_j^2 + \frac{1}{2\sigma_e^2}\sum_i e_i^2 + W \quad \text{s.t.}\quad p(X\beta + e) \leq pt \quad (11)$$

where $W$ is a quantity that does not depend on $\beta$ or $e$, and can thus be ignored. This is a standard quadratic optimization problem, amenable to exact solution using standard convex optimization techniques[45]. Given the joint MAP of $\beta$ and $e$, $\hat{l}$ is given by $\hat{l} = X\hat{\beta} + \hat{e}$.

### *Genetic MAP Estimator*

The MAP of the effect sizes $\beta$ can be found by maximizing their posterior likelihood. Using the same derivation as before, with the exception that $e$ is integrated out, the quantity to maximize is given by

$$\varphi\left(\beta;\ 0,\frac{\sigma_g^2}{m}I\right) \prod_{i \in controls} \Phi(t - X_i^T\beta;\ 0,\sigma_e^2) \prod_{i \in cases}\left(1 - \Phi(t - X_i^T\beta;\ 0,\sigma_e^2)\right). \quad (12)$$

Taking the logarithm and using the normal distribution definition, the quantity to maximize is

$$\sum_{i \in controls} \log \Phi(t - X_i^T \beta; 0, \sigma_e^2) + \sum_{i \in cases} \log \left(1 - \Phi(t - X_i^T \beta; 0, \sigma_e^2)\right)$$
$$- \frac{1}{2\sigma_g^2/m} \sum_j \beta_j^2 + W \quad (13)$$

where $W$ is a quantity that does not depend on $\beta$ and can thus be ignored. This problem is equivalent to Probit regression[23] with L2 regularization and a pre-specified offset term, and can thus be solved using standard techniques (Supplementary Note 2). Unlike typical uses of such models, here the regularization parameter is known in advance, given a value for $\sigma_g^2$.

The MAP $\hat{g}$ is given by $\hat{g} = X\hat{\beta}$, where $\hat{\beta}$ is the MAP of $\beta$. Given the MAP $\hat{g}$, $\hat{l}$ is determined by setting the entries of all cases with $\hat{g}_\iota < t$, and all controls with $\hat{g}_\iota > t$, to be equal to $t$. All other entries in $\hat{l}$ are equal to the corresponding entry in $\hat{g}$. This follows because $e$ has a zero-mean normal distribution.

We opted to use the genetic MAP estimator, rather than the joint MAP estimator, because it is more suitable for liability estimation in the presence of related individuals. This greater suitability comes from the way LEAP handles related individuals, which consists of first excluding them from the model fitting stage, and then estimating their liabilities via the fitted model (see further details below). The joint MAP estimator minimizes the in-sample estimation error of the liabilities, because it directly fits the environmental component $e$ of individuals participating in the fitting stage. In contrast, the genetic MAP estimator attempts to minimize the out-of-sample estimation error, because it integrates the environmental component $e$ out and only fits the effect sizes $\beta$. The effect sizes $\beta$ are later used to estimate liabilities for all individuals, including those that did not participate in the model fitting stage. Therefore, the genetic MAP estimator is more suitable for the purposes of LEAP.

We verified empirically that the genetic MAP estimator often yields more accurate estimates than either the MAP or the posterior mean estimator (Supplementary Note 1, Supplemental Fig. S20 and Supplementary Table S4).

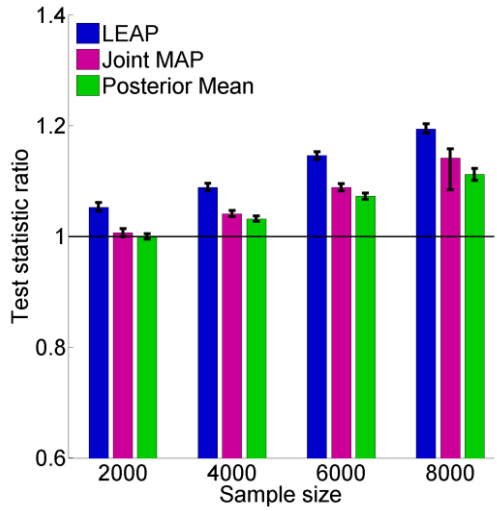

**Supplementary Figure S20**: Comparison of LEAP with the results of GWASs performed with different liability estimators: The joint MAP estimator and the posterior mean estimator, evaluated via sampling. The results show the ratio between the normalized test statistics of each method and a standard LMM. The posterior mean estimator used the average of the true frequency of each SNP (between the two populations) for covariance estimation, as described in Supplementary Note 1.